\documentclass[acmsmall,screen,nonacm]{acmart}
\usepackage{graphicx} 
\usepackage{amsmath}
\usepackage{multirow}
\usepackage{float}
\usepackage{subfig}
\usepackage{booktabs}
\usepackage[htt]{hyphenat}
\usepackage{makecell}
\usepackage{tabularx}
\usepackage{enumitem}
\usepackage{wrapstuff}
\makeatletter
\newcount\aw@cnt \newcount\aw@lines
\NewDocumentEnvironment{autowrap}{m m O{} +b}{%
  \par
  \if@nobreak\else\penalty\z@\fi
  \setbox\z@\vbox{\aw@cnt\value{#1}\let\label\@gobble\def\@captype{#1}%
    \begin{minipage}{#2}#4\end{minipage}\setcounter{#1}{\aw@cnt}}%
  \dimen@\pagegoal \advance\dimen@-\pagetotal
  \aw@lines\z@
  \ifdim\pagegoal<\maxdimen
    \ifdim\dimexpr\ht\z@+\dp\z@+0.5\baselineskip\relax>\dimen@
      \ifdim\dimen@>\baselineskip
        \aw@lines\numexpr(\dimen@-\baselineskip/2)/\baselineskip\relax
      \fi
    \fi
  \fi
  \def\aw@type{#1}\def\aw@alg{algorithm}%
  \ifnum\aw@lines>\z@
    \ifx\aw@type\aw@alg
      \begin{algorithm}[t]#4\end{algorithm}%
    \else
      \begin{wrapstuff}[r,type=#1,width=#2,top=\the\aw@lines,#3]#4\end{wrapstuff}%
    \fi
  \else
    \begin{wrapstuff}[r,type=#1,width=#2,#3]#4\end{wrapstuff}%
  \fi}{}
\newcommand{\placehereornext}[1]{%
  \par
  \setbox\z@\vbox{\aw@cnt\value{figure}\let\label\@gobble\def\@captype{figure}%
    #1\setcounter{figure}{\aw@cnt}}%
  \dimen@\pagegoal \advance\dimen@-\pagetotal
  \ifdim\dimexpr\ht\z@+\dp\z@+\intextsep\relax>\dimen@
    \afterpage{\begin{figure}[t]#1\end{figure}}%
  \else
    \begin{figure}[H]#1\end{figure}%
  \fi}
\makeatother
\usepackage{algorithm}
\usepackage{afterpage}
\usepackage{algpseudocode}
\usepackage[most]{tcolorbox}
\usepackage{xhfill}
\newtcolorbox{finding}[1]{%
  enhanced, breakable,
  colback=gray!6, colframe=black!60,
  boxrule=0.5pt, arc=1pt, boxsep=2pt,
  left=3pt, right=3pt, top=2pt, bottom=2pt,
  before skip=5pt, after skip=5pt,
  fontupper=\small,
  before upper={\textbf{Finding #1.}\ }}
\newtcolorbox{answer}[1]{%
  enhanced, breakable,
  colback=gray!6, colframe=black!60,
  boxrule=0.5pt, arc=1pt, boxsep=2pt,
  left=3pt, right=3pt, top=3pt, bottom=3pt,
  before skip=5pt, after skip=5pt,
  fontupper=\small,
  before upper={\textbf{Answer to RQ#1.}\ }}

\setcopyright{none}
\usepackage{xspace}
\newcommand{\Exact}{\texttt{Exact}\xspace}
\newcommand{\Partial}{\texttt{Partial}\xspace}

\definecolor{nwInk}{HTML}{1A1F24}
\definecolor{nwLabel}{HTML}{5B6672}
\definecolor{nwFrame}{HTML}{8A94A0}
\definecolor{nwBlue}{HTML}{2E4AA8}
\definecolor{nwRed}{HTML}{B3412E}
\definecolor{nwAmber}{HTML}{D9A441}
\definecolor{nwAmberBg}{HTML}{FFF3D6}

\newcommand{\caseanswer}[3]{\colorbox{#1}{\parbox{\dimexpr\linewidth-2\fboxsep\relax}{%
    \textit{Answer:} \texttt{#2}\hfill\textit{#3}}}}

\begin{document}
\title{Fault-Free Profiles and Decision-Level Training for Root Cause Analysis via Small Language Models}

\author{Kyoohyun Lee}
\email{khyhlee@hanyang.ac.kr}
\affiliation{%
\department{Department of Artificial Intelligence}
  \institution{Hanyang University}
  \city{Seoul}
  \country{Republic of Korea}
}

\author{Woohyeok Park}
\email{woohyeok@hanyang.ac.kr}
\affiliation{%
  \department{Department of Data Science}
  \institution{Hanyang University}
  \city{Seoul}
  \country{Republic of Korea}
}

\author{Taeyoon Kim}
\email{tykim7@hanyang.ac.kr}
\affiliation{%
  \department{Department of Data Science}
  \institution{Hanyang University}
  \city{Seoul}
  \country{Republic of Korea}
}

\author{Sungyong Baik}
\email{dsybaik@hanyang.ac.kr}
\affiliation{%
  \institution{Hanyang University}
  \city{Seoul}
  \country{Republic of Korea}
}

\author{Kyungyong Lee}
\email{kyungyong@hanyang.ac.kr}
\affiliation{
   \department{Department of Data Science}
  \institution{Hanyang University}
  \city{Seoul}
  \country{Republic of Korea}
}

\begin{abstract}
Root cause analysis (RCA) in microservice systems requires distinguishing failure sources from routine fluctuations and propagated symptoms. Small language models (SLMs) offer a practical solution under cost and privacy constraints, and recent approaches improve their capabilities through structured diagnostic procedures. However, performing diagnostic analysis does not ensure correct root-cause identification. Our empirical study shows that final decisions often favor large anomaly scores despite the model's own propagation analysis suggesting otherwise. Moreover, the input does not reveal whether these apparently anomalous signals are common during normal operation. These findings motivate our system, which improves how SLMs assess evidence and select root causes through Historical Frequency Profiling and Biased-Decision Correction. Historical Frequency Profiling quantifies how often metric and log signals are flagged as anomalous in the fault-free period, providing context for distinguishing routine deviations from potential failure evidence. 
Biased-Decision Correction identifies points where the model reaches different conclusions from the same initial analysis and uses preference optimization to favor more accurate diagnoses. Against baselines including the 14B OpsAgent and RCA-Agent with commercial models, our system with a 4B model achieves the highest or tied-highest accuracy on all four datasets, reaching 48.7\,\% on Bank compared with 18.5\,\% for OpsAgent.
\end{abstract}

\maketitle

\section{Introduction}\label{sec:intro}
Modern cloud services consist of interdependent microservices, where a fault in one component can produce symptoms across many others~\cite{icse26-foundroot, stalled-forge26, zhou2021trainticket}. Root Cause Analysis (RCA) distinguishes the failure source from propagated symptoms and identifies the faulty component, fault type, and occurrence time~\cite{openrca}. This requires interpreting telemetry of metrics, logs, and traces, and its volume and complexity quickly exceed what on-call engineers can manually inspect.

Classical RCA methods analyze statistical dependencies and event patterns to localize failure sources~\cite{iwqos20-microcause, icse21-microhecl, fse23-nezha}. Recent work has shifted attention to large language models (LLMs), which interpret telemetry and explain how observations support a diagnosis~\cite{openrca, icse26-foundroot, ase-opsagent}. To fit telemetry into a model's limited context window, RCA systems typically retain signals flagged as anomalous and summarize them as a text prompt, a step we refer to as \emph{data processing}~\cite{ase-opsagent, openrca}. This step determines the evidence available to the model, which examines it alongside component dependencies to identify the root cause.

However, API costs and restrictions on sharing operational telemetry motivate self-hosted small language models (SLMs)~\cite{ase-opsagent, tosem-thinkfl, rcacopilot-eurosys24}. Their diagnostic accuracy remains limited. Given prompts built by existing data processing, the off-the-shelf SLMs we evaluate correctly answer all requested elements for fewer than 4\,\% of incidents. Frontier models achieve nearly three times the accuracy on the same prompts, at substantially higher inference costs (\S\ref{sec:llm-vs-slm}).

Recent approaches improve diagnosis by organizing how models investigate incidents. RCA-Agent queries telemetry through tools~\cite{openrca}, while SLM-oriented systems such as OpsAgent~\cite{ase-opsagent}, FoundRoot~\cite{icse26-foundroot}, and ThinkFL~\cite{tosem-thinkfl} guide or train models to examine observations, analyze propagation, and refine diagnoses. However, a model that follows such a procedure can still select the wrong root cause. It may note that a large anomaly could be routine behavior or a symptom propagated from another component, yet still choose the component showing that anomaly.

We investigate this gap through an empirical study of data processing and Qwen3-4B's diagnostic decisions on the Bank dataset of OpenRCA~\cite{openrca}, revealing two limitations.

\emph{The input lacks normal-behavior context.} The examined data processing flags signals with short-term statistics and fixed thresholds, so fluctuations during normal operation are flagged as well. For example, garbage collection causes a JVM memory metric on Bank to be flagged in 97\,\% of fault-free windows. Such signals still reach the prompt as failure evidence, with large anomaly scores and no indication of how common they are. As a result, the actual root-cause component has the largest anomaly score in only 8\,\% of incident windows, even though it is among the candidates in 99\,\% (\S\ref{sec:study-order}). 

\emph{The final decision often departs from the model's own analysis.} Qwen3-4B examines multiple candidates and propagation relationships in all 640 trajectories, yet it justifies its selection by large anomaly scores in 76\,\% of them and departs from its own propagation analysis in 68\,\%. Adding a few-shot example with a correct diagnosis of another incident does not reliably fix this behavior. The gains depend largely on whether the example names the same root-cause component as the target incident (\S\ref{sec:empirical-study-procedure}).

These findings reveal a gap between performing diagnostic analysis and selecting the root cause. Building on this insight, we present an SLM-based RCA system that addresses this gap by providing normal-behavior context for assessing anomalies and targeting decisions where the model reaches different conclusions from the same initial analysis.

\textbf{Historical Frequency Profiling} measures how often metric and log signals are flagged as anomalous in the fault-free period. This frequency, called the Expected Outlier Baseline (EOB), accompanies anomaly scores and guides candidate ordering. Traces are summarized by their latency relative to that in the fault-free period. These profiles help distinguish routine deviations from potential failure evidence without treating rarity alone as proof of a root cause.

\textbf{Biased-Decision Correction} compares the model's own responses that share the same initial analysis but reach different diagnostic conclusions. Preference pairs reflect correctness on individual RCA elements and favor more complete diagnoses among partially correct responses. When none of the responses is correct, auxiliary hints help obtain a correct one, which is checked against the original evidence. The hints are removed before training. Direct Preference Optimization (DPO)~\cite{neurips23-dpo} then favors more accurate responses by training on the reasoning and decisions following the shared initial analysis.

We evaluate our system on eight datasets. Telecom, Bank, and Market from OpenRCA~\cite{openrca} and AIOps 2025~\cite{aiops2025} are used for evaluation within each system, and four held-out datasets are used to test whether an adapter trained on one system transfers to unseen systems. With a 4B model and one inference call per incident, our system achieves the highest exact-match accuracy on Telecom, Bank, and AIOps 2025 and ties with the best baseline on Market. On Bank, it reaches 48.7\,\%, compared with 18.5\,\% for OpsAgent-14B~\cite{ase-opsagent}. Under identical input, Biased-Decision Correction also enables the 4B model to outperform an off-the-shelf 14B model. A Bank-trained adapter exceeds FoundRoot-14B in exact-match accuracy on the three held-out datasets with full diagnostic labels, using target-system fault-free profiles without further fine-tuning.

Table~\ref{tab:related} (\S\ref{sec:related}) compares our system with existing approaches in input evidence, diagnostic scope, and training target. In summary, this paper makes the following contributions.
\begin{itemize}[leftmargin=*]
    \item An empirical study revealing missing normal-behavior context and a gap between the evaluated SLM's diagnostic analysis and final root-cause selection.
    \item An SLM-based RCA system that contextualizes anomaly evidence with fault-free frequencies and targets diagnostic decisions that diverge after the same initial analysis.
    \item An evaluation across eight datasets demonstrating improved accuracy with a 4B model, including comparisons under identical input and transfer without further fine-tuning.
\end{itemize}

\section{Root Cause Analysis in Microservice Systems}\label{sec:rca}

\begin{figure}[t]
  \centering
  \subfloat[Call graph of Market\label{fig:bg-callgraph}]{%
    \includegraphics[width=0.31\linewidth]{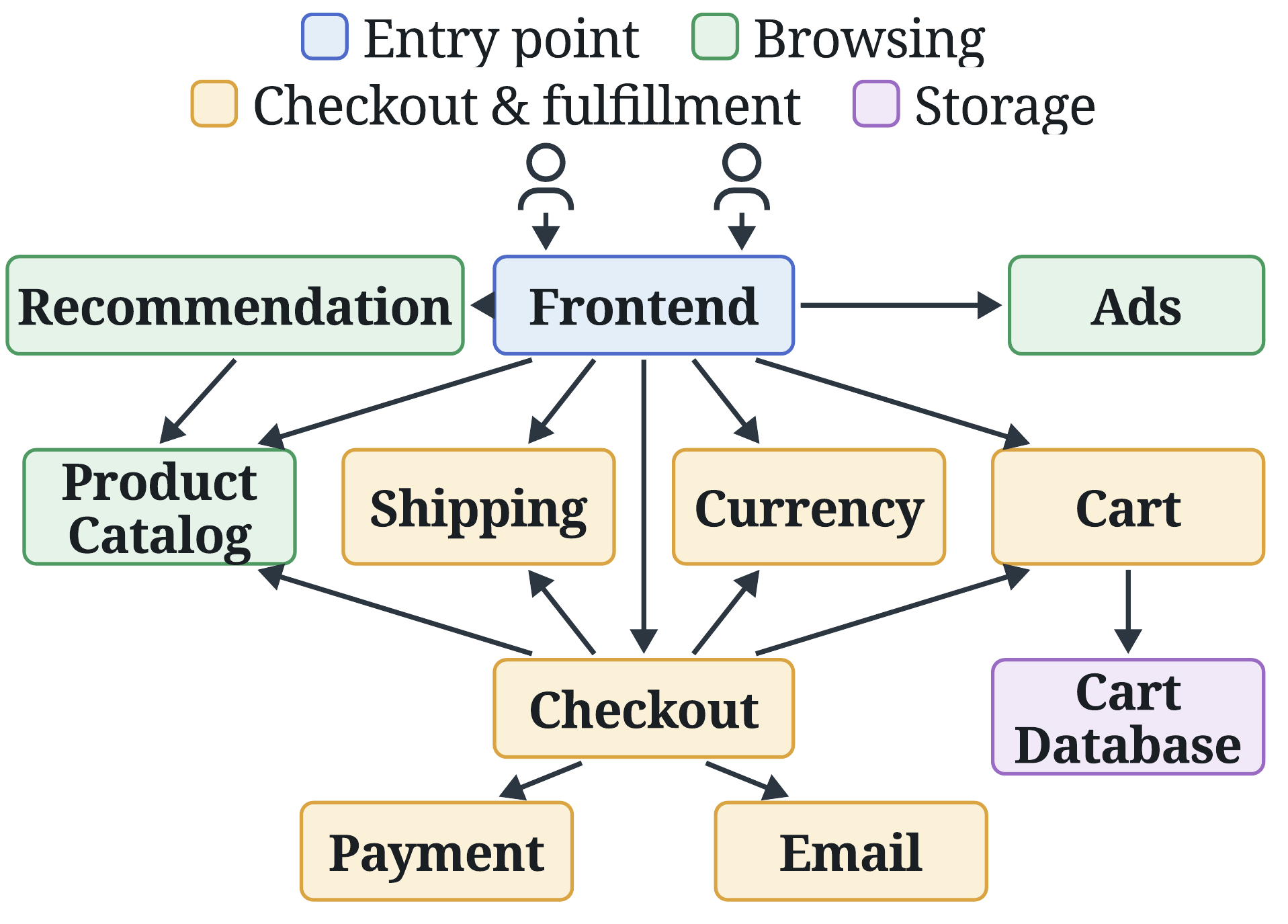}%
  }%
  \hfill
  \subfloat[Telemetry of \texttt{shippingservice-1} in Market during an incident\label{fig:bg-telemetry}]{%
    \includegraphics[width=0.64\linewidth]{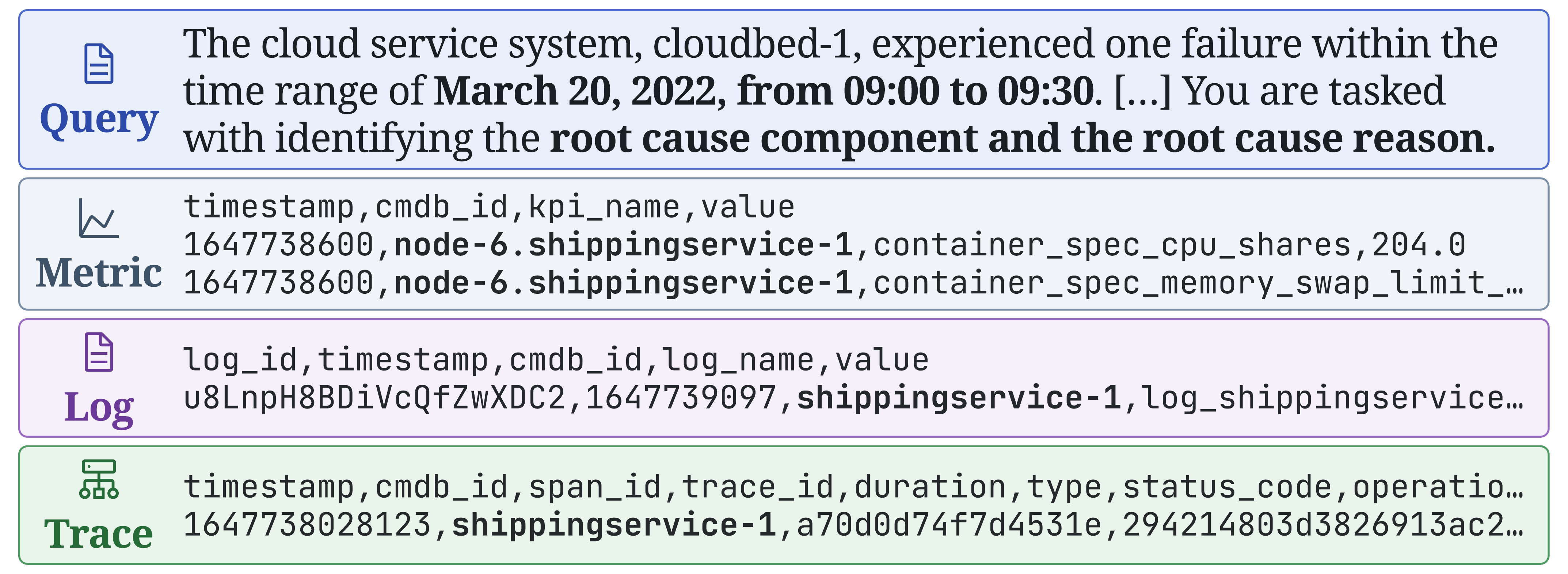}%
  }
  \caption{\textbf{Inputs of microservice RCA in OpenRCA~\cite{openrca}.}}
  \label{fig:system}
\end{figure}

A microservice system serves each user request through a chain of services, so a fault in one component propagates to the components that depend on it. In the Market system of OpenRCA~\cite{openrca} (Fig.~\ref{fig:bg-callgraph}), for example, a slow response of Shipping is also observed at Checkout and Frontend. RCA must therefore distinguish the root-cause component from the components that only show propagated symptoms~\cite{icse26-foundroot, tosem-thinkfl, stalled-forge26}.
 
RCA relies on three types of telemetry. Metrics are time series recorded periodically, such as CPU utilization, logs are semi-structured text records emitted by components, and traces record each service call of a request as a span with its start time, duration, and status code~\cite{openrca}. Fig.~\ref{fig:bg-telemetry} shows the three types for \texttt{shippingservice-1}, an instance of Shipping, during an incident. Because they are recorded separately, diagnosis must link them by time and component.

\subsection{Problem Formulation}\label{sec:formulation}
A microservice system consists of a set of components $\mathcal{C}$, such as hosts, instances, and services (e.g., \texttt{node-6} and \texttt{shippingservice-1} in Fig.~\ref{fig:bg-telemetry}), and a call graph $G$ among them, which is obtained from deployment information or from traces aggregated over a long period. An on-call engineer sets an incident window $W = [t_0, t_1]$ as the period in which the failure is suspected~\cite{ase-opsagent}. The telemetry observed in $W$ is denoted by $\mathcal{T}_W = (\mathcal{M}_W, \mathcal{L}_W, \mathcal{R}_W)$, the sets of metric time series, log entries, and spans in $W$, respectively. Because $\mathcal{R}_W$ records only the calls made during $W$, dependencies between components that were not called in $W$ are available only through $G$.

A fault of type $f \in \mathcal{F}$ that occurs at component $c \in \mathcal{C}$ at time $t$ is described by the label $y = (c, f, t)$, where $\mathcal{F}$ is a predefined set of fault types such as high CPU usage and network packet loss. An \emph{incident} consists of one or more faults within an incident window, with a query $q$ that asks for some of these three elements. RCA predicts the requested elements from $q$, $\mathcal{T}_W$, and $G$ as
\begin{equation}
  \hat{y} = (\hat{c}, \hat{f}, \hat{t}) = \Psi(q, \mathcal{T}_W, G),
  \quad \hat{c} \in \mathcal{C},\; \hat{f} \in \mathcal{F},\; \hat{t} \in W.
  \label{eq:rca}
\end{equation}

\subsection{LLM-based RCA Approaches}\label{sec:bg-approaches}
Classical RCA methods detect anomalous signals, relate them over the call graph, and rank candidate components, using either statistical scores~\cite{icse21-microhecl, iwqos20-microcause, ccgrid18-cloudranger, icse24-tracecontrast, fse22-tracecrl, fse19-latenterror} or models trained on past incidents~\cite{fse22-dejavu, icse23-eadro}. Because these methods rank signals by statistical patterns rather than by what a metric, log, or trace means, metric noise and complex fault propagation can mislead their ranking~\cite{icse26-foundroot}. They also output a ranked list of candidates rather than a diagnosis that relates the evidence to the root cause~\cite{ase-opsagent, tosem-thinkfl}.
 
LLM-based RCA systems address these limitations~\cite{icse23-ahmed, rcacopilot-eurosys24, fse24-rcaagents, openrca, icse26-foundroot, ase-opsagent, zhou2026tracellm}. Drawing on the domain knowledge acquired during pre-training, an LLM interprets signals, infers fault propagation along the call graph, and explains its conclusion in natural language, for example through chain-of-thought (CoT) reasoning~\cite{cot}. Because raw telemetry exceeds the context window~\cite{openrca, tosem-thinkfl, tamo}, these systems condense telemetry into a prompt through data processing and guide the model with a diagnostic procedure, given either in the prompt or through training.
 
\subsubsection{Data Processing: Curating Inputs by Anomaly Statistics}\label{sec:bg-curation}
Data processing, which selects anomalous signals statistically and converts them into a prompt~\cite{ase-opsagent, icse26-foundroot, openrca, tosem-thinkfl, arxiv25-microrcaagent}, determines the scope and format of the model input. It first divides telemetry into signals. A metric signal is a time series identified by a component and a key performance indicator (KPI) such as CPU utilization. A log signal is the set of entries of one template emitted by a component, where a template, extracted by a log parser such as Drain~\cite{drain3}, is the fixed text of an entry without variable parts such as identifiers and numeric values~\cite{jiang2024lilac, huang2025lunar}. A trace signal is the set of spans of all calls to a component. We denote the set of signals in $W$ by $\mathcal{S}_W$.
 
Data processing then assigns each signal $s \in \mathcal{S}_W$ an anomaly score $\phi(s)$ and keeps only the signals whose score exceeds a threshold $\tau$, yielding $\tilde{\mathcal{S}}_W = \{ s \in \mathcal{S}_W : \phi(s) > \tau \}$~\cite{ase-opsagent}. For a metric signal, $\phi$ is the z-score $|v - \mu| / \sigma$ at the point in $W$ where the observed value $v$ deviates the most from the mean. $\mu$ and $\sigma$ are estimated from a short comparison interval, typically the few minutes around $W$, and $\tau$ is typically 3 (the 3-sigma rule)~\cite{ase-opsagent, icse26-foundroot}. For a log signal, $\phi$ is the number of entries of the template that the component emits in $W$, counted only for templates that contain keywords such as fatal, error, and crash~\cite{ase-opsagent}. For a trace signal, $\phi$ is the number of spans whose latency exceeds a threshold set for their call type~\cite{ase-opsagent}. For logs and traces, $\tau$ is 0, so a signal remains if any entry satisfies the condition.

The selected signals are then grouped by component, and components with at least one selected signal form the candidate set $\mathcal{E}_W$. Data processing then writes $\mathcal{E}_W$, the selected signals of each candidate with their $\phi$, and $G$ as text with $\mathrm{render}$ to build the prompt $p$, which also contains the query and output format~\cite{icse26-foundroot, ase-opsagent}. LLM-based RCA thus splits $\Psi$ in Eq.~\eqref{eq:rca} into two steps: data processing builds $p$ from $q$, $\mathcal{T}_W$, and $G$, and the model outputs $\hat{y}$ from $p$. The model can therefore choose only from $\mathcal{E}_W$ and judge each candidate only by its selected signals and their $\phi$: a root cause missing from $\mathcal{E}_W$ cannot be found, and a misleading $\phi$ can draw the model to a wrong candidate.

\subsubsection{In-Context Learning}\label{sec:bg-procedure}
In-context learning changes the behavior of a model by adding instructions or examples $d$ to the prompt while keeping its weights fixed~\cite{incontextlearning}. The model receives $p \oplus d$, where $\oplus$ denotes the concatenation of two token sequences. Without such guidance, the model reaches its conclusion through a different process in each trial, for example by skipping some components or by accepting the first anomaly it finds as the cause~\cite{icse26-foundroot, stalled-forge26, kim2026aiagentssystematicallyfail}. A pre-trained model also has general domain knowledge but lacks knowledge specific to the target system, such as the diagnostic steps that its operators follow~\cite{ase-opsagent, rcacopilot-eurosys24}.
 
LLM-based RCA systems therefore provide the model with a \emph{diagnostic procedure} that specifies which components to examine and in what order, such as scanning metrics before analyzing propagation along the call graph. The procedure is either stated as instructions~\cite{stalled-forge26, cot} or given as the diagnosis records of similar past incidents~\cite{rcacopilot-eurosys24, ase-opsagent}.

\subsubsection{Fine-Tuning: SFT and RL}\label{sec:bg-finetune}
Fine-tuning updates the weights of a model with examples or rewards~\cite{fine-tuning}. Supervised fine-tuning (SFT) maximizes the likelihood of a target output for each input, and reinforcement learning (RL) maximizes the expected reward of the outputs that the model generates, with algorithms such as PPO~\cite{ppo}, GRPO~\cite{grpo}, and DAPO~\cite{arxiv25-dapo}. Direct Preference Optimization (DPO)~\cite{neurips23-dpo} instead trains the model on pairs of preferred and dispreferred responses to the same input, without a separate reward model.
 
For RCA, prior work performs SFT on trajectories that follow a predefined diagnostic procedure and then improves accuracy with RL~\cite{tosem-thinkfl, icse26-foundroot, ase-opsagent}. Both methods apply the training signal to every token of an output, so they adjust the diagnostic procedure and the final decision together.

\subsection{Frontier Models and SLMs for RCA}\label{sec:llm-vs-slm}
Deploying LLM-based RCA requires choosing between frontier models accessed through APIs and SLMs that can be self-hosted. We refer to open-weight models with up to 20B parameters that can be served on a single GPU as SLMs. To compare the two, we use the accuracy measure of OpenRCA~\cite{openrca}, the fraction of incidents for which all elements, such as component, fault type, and occurrence time, asked by the query are correct. We call this measure exact-match accuracy and denote it by \Exact.

We ran both types of models on three OpenRCA datasets (Telecom, Bank, and Market) through Amazon Bedrock~\cite{bedrock}, giving all models the same prompt built by the data processing of OpsAgent~\cite{ase-opsagent}. Claude Opus 5~\cite{claude} and GPT-5.6 Sol reached an \Exact of 10.5\,\% and 10.6\,\% at \$42.1 and \$28.5 per 1,000 incidents, respectively. Off-the-shelf Gemma 3 4B, Gemma 3 12B, and gpt-oss-20b cost only \$0.2--0.4 per 1,000 incidents but reached only 3.7--3.9\,\%.
 
Frontier models are thus more accurate, but their API costs grow with the number of incidents, and they require sending operational data outside the organization~\cite{rcacopilot-eurosys24, ase-opsagent}. Self-hosted SLMs avoid both problems, so improving their accuracy is the practical path for RCA in production. This requires locating where accuracy is lost, either in the prompt built by data processing or in the cause selection of the SLM, which we examine in the next section.

\section{Empirical Study}\label{sec:empirical-study}
The accuracy of LLM-based RCA depends on both the input built by data processing~\cite{openrca, ase-opsagent, tamo, stalled-forge26} and the cause selection of the model on that input~\cite{openrca, icse26-foundroot, stalled-forge26}. We examine the two separately through the following questions.

\begin{itemize}[leftmargin=*]
  \item \textbf{Q1.} Does the input $p$ built by existing data processing contain information that distinguishes the root-cause component from other candidates? (\S\ref{sec:study-order})
  \item \textbf{Q2.} What limits the accuracy of the SLM in selecting the root cause? (\S\ref{sec:empirical-study-procedure})
\end{itemize}

We study both questions on the Bank dataset of OpenRCA~\cite{openrca}, collected from a banking microservice system. It contains 10 days of metrics, logs, and traces and 136 incidents from injected faults, each with a query that specifies a 30-minute incident window. Bank labels the fault type of every fault and has relatively few components, which suits qualitative analysis.

\subsection{Prompts Built by Existing Data Processing}\label{sec:study-order}
We analyze the prompt built by the data processing of OpsAgent~\cite{ase-opsagent}, which contains the anomaly scores $\phi(s)$ (z-scores) of metrics, selected log entries, and summaries of high-latency spans. For metrics, $\mu$ and $\sigma$ are estimated from a short comparison interval before the incident window $W$, as in existing data processing~\cite{ase-opsagent, icse26-foundroot}, and a signal is flagged as anomalous when $\phi(s)$ exceeds $\tau = 3$.

To check whether $\phi$ distinguishes the root cause, we measure how often each signal is flagged as anomalous when no fault is present. We call 
the time outside all incident windows the \emph{fault-free period}~\cite{fse23-nezha}. Excluding 5 minutes before and after each incident window as a margin, we split this period into 10-minute \emph{fault-free windows} and apply the same data processing to each of them.

\begin{autowrap}{figure}{0.48\linewidth}
\centering
\includegraphics[width=\linewidth]{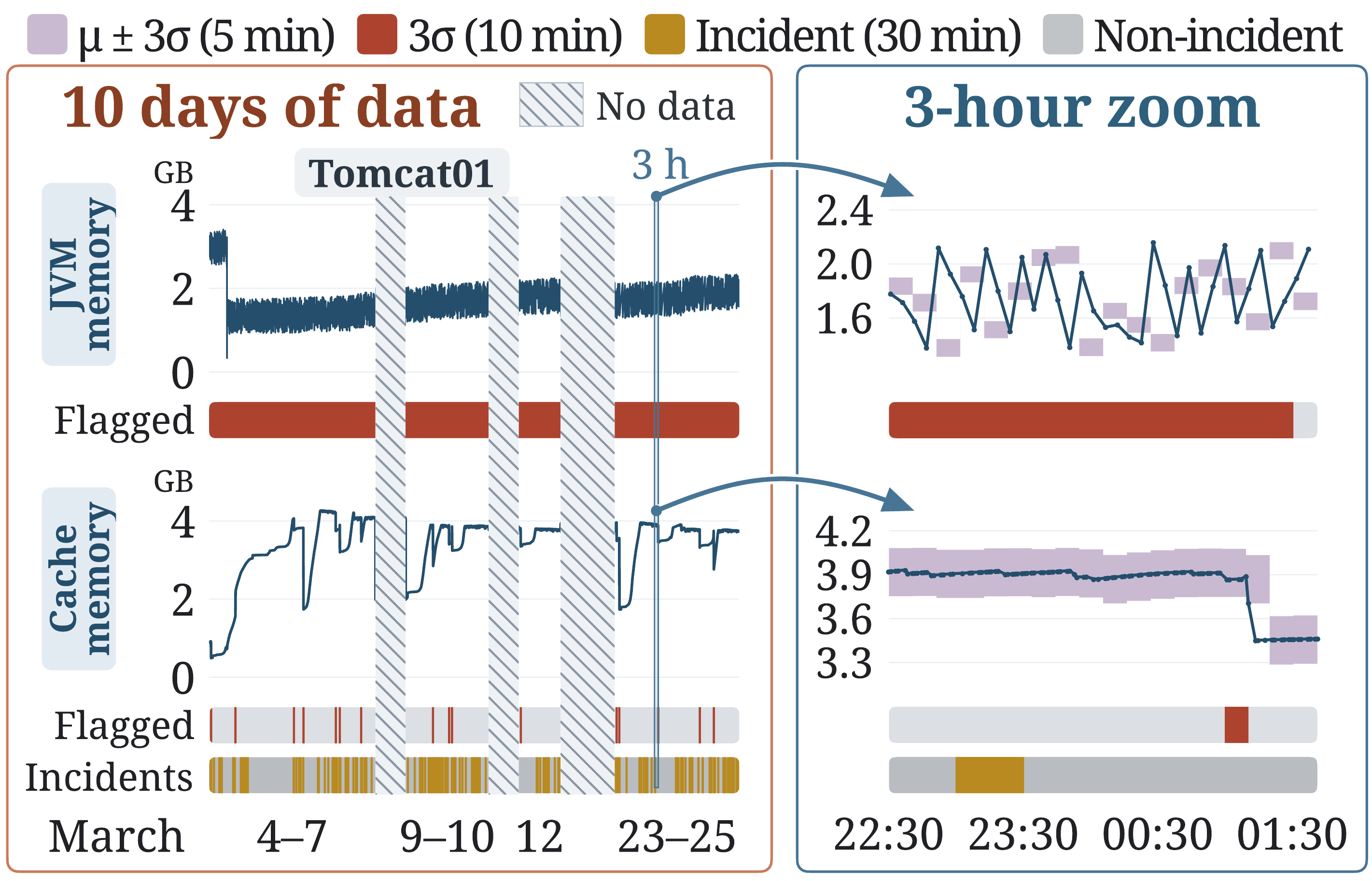}
\caption{\textbf{Signals exceeding $\tau$ in fault-free windows vs.\ only during failures.} }
\label{fig:3-normal-window}
\end{autowrap}

Fig.~\ref{fig:3-normal-window} shows two metrics of Tomcat01. JVM memory rises and falls every 5 minutes with garbage collection (GC) and exceeds $\tau$ in 97\,\% of the fault-free windows. In contrast, cache memory exceeds $\tau$ in only 17 windows over 10 days, and 13 of them fall within incident windows. When cache memory exceeds $\tau$ in $W$, a fault is therefore the likely cause, yet the prompt presents both signals in the same format without showing how often each exceeds $\tau$ in the fault-free period. JVM memory is misjudged because the 5 minutes before $W$ miss the full range of its GC cycle, which underestimates $\sigma$. With $\mu$ and $\sigma$ from the whole fault-free period instead of the preceding 5 minutes of the incident window, the false positive rate drops from 97\,\% to 4\,\%.

Logs and traces show the same pattern. For logs, data processing flags entries that contain keywords such as error and failure, and the JVM writes a GC entry such as ``\texttt{[GC (Allocation Failure)...]}'' for every routine young-generation collection. The GC logs of Tomcat01--04 are therefore flagged as anomalous in 82--87\,\% of the fault-free windows, but the Tomcat that writes them is the root cause in only 8--15\,\% of the incident windows where they appear. For traces, data processing counts the high-latency spans of the calls each component receives, such as ``\texttt{Callee=IG01, Count=110, MaxLatency=10865ms}''. Because its released implementation sets the latency threshold from the spans in the incident window itself, some spans exceed it in every window. The spans of IG01 and IG02 are flagged as high latency in 97--98\,\% of the fault-free windows, but IG01 or IG02 is the root cause in only 7--14\,\% of the incident windows. Such signals act as false positives~\cite{fse23-nezha, icse20-alertstorm}, but they make up most of the log and trace evidence in the prompts. As a result, although the root-cause component is in $\mathcal{E}_W$ for 99\,\% of incident windows, it has the largest $\phi$ for only 8\,\%.

\begin{finding}{1}
Existing data processing compares each signal against the incident window or the few minutes around it, so the prompt does not show how each signal behaves without faults. In Bank, the root cause is a candidate in 99\,\% of incident windows but has the largest $\phi$ in only 8\,\%. Exceeding $\tau$ in $W$ alone does not point to the root cause, and the prompt should also state how often each signal exceeds $\tau$ without faults.
\end{finding}

\subsection{Root Cause Selection by the SLM}\label{sec:empirical-study-procedure}

Prior work provides a diagnostic procedure in the prompt or trains the model to follow it~\cite{icse26-foundroot, ase-opsagent, tosem-thinkfl}. We examine whether the SLM performs the diagnostic procedure by itself and whether providing the procedure reduces errors in cause selection. For this analysis, we collected 640 reasoning trajectories that Qwen3-4B~\cite{qwen3} generated, eight per incident, from the prompt described above for the 80 Bank incidents in our training split. We then analyzed how candidate examination and intermediate reasoning lead to the final answer.

We labeled each trajectory with two independent judgments by Claude Opus 5 as an LLM judge~\cite{llm-as-judge} and resolved disagreements with a third. The per-item agreement between the two judgments was 88--100\,\%, and Cohen's $\kappa$ was 0.68--0.83. The authors then reviewed all final judgments. In all trajectories, the model examined the signals of two or more candidates and the propagation relationships between components. When selecting the cause, however, it repeatedly showed three behaviors that kept its final answer from following this analysis.

\begin{figure}[t]
  {\scriptsize\color{nwInk}\setlength{\fboxsep}{2pt}\setlength{\fboxrule}{0.6pt}%
  \newcommand{\steplabel}[1]{\textsf{\textbf{\textcolor{nwLabel}{#1}}}\enspace}%
  \newcommand{\bw}{\dimexpr\linewidth-2\fboxsep-2\fboxrule\relax}%
  \newcommand{\trstep}[2]{\noindent\fcolorbox{nwFrame}{white}{\parbox{\bw}{\steplabel{#1}#2}}\par\vspace{2pt}}%
  \newcommand{\trans}[2]{\noindent\fcolorbox{nwAmber}{nwAmberBg}{\parbox{\bw}{\steplabel{#1}#2}}\par}%
  \newcommand{\good}[1]{\textcolor{nwBlue}{\textbf{#1}}}%
  \newcommand{\bad}[1]{\textcolor{nwRed}{\textbf{#1}}}%
  \newcommand{\trcase}[3]{\noindent\textbf{Case} #1.\quad GT: \good{#2}.\quad Answer: \bad{#3}.\par\vspace{2pt}}%
  \subfloat[A final decision that contradicts the intermediate conclusion\label{fig:behavior-a}]{\parbox{\linewidth}{%
  \trcase{openrca\_bank\_1616667180\_107, trajectory 7}{IG02}{Tomcat01}
  \trstep{Scan}{Looking at the metrics, Tomcat01, Tomcat02, Tomcat03, and Tomcat04 have several issues. For example, Tomcat01 has high JVM memory usage and disk I/O spikes. [\ldots] \good{IG01 and IG02 have local disk I/O problems, like high DSKBps and DSKWrite.}[\ldots]}
  \trstep{Propagation}{Looking at the component graph, Tomcat instances depend on MG (MG01, MG02), Redis01, Redis02, and the OS resources. The propagation shows that Tomcat01, 02, 03, 04 are downstream of IG01, IG02, which are downstream of apache01 and 02. \good{So the problem might be starting from the apache servers, which are upstream of the Tomcats.}}
  \trans{Ranking}{So the top components would be Tomcat instances, as they have the most direct issues with memory and CPU. Then the disk metrics for IG01 and IG02, which are upstream of Tomcat. Finally, the MG01 and MG02, which are dependencies of Tomcat [\ldots] \textbf{Answer:} \bad{1. Tomcat01}: High JVM memory usage and frequent GC errors indicate memory pressure leading to crashes or slowdowns. 2. IG01. 3. MG01.}
  }}\par\vspace{4pt}
  \subfloat[Misattributing a signal of another component\label{fig:behavior-b}]{\parbox{\linewidth}{%
  \trcase{openrca\_bank\_1614866400\_50, trajectory 5}{MG01 (JVM OOM heap)}{Tomcat01}
  \trstep{Input}{Tomcat01: CurrentThreadsBusy ($z{=}30$), sdb\_DSKBps ($z{=}30$), CPU-2 idle ($z{=}30$) [\ldots] \good{IG01: JVM-Memory\_7778\_HeapMemoryUsed ($z{=}6.0$, max 1017709120)}, JVM-Memory\_7778\_HeapMemoryUsage ($z{=}5.5$), sda\_DSKWTps ($z{=}6.1$) [\ldots]}
  \trstep{Reasoning}{Looking at the component graph, Tomcat01, Tomcat02, Tomcat03, Tomcat04 are dependent on IG01 and IG02, which are upstream [\ldots] Also, the metrics for \bad{Tomcat01's JVM memory usage (JVM-Memory\_7778\_JVM\_Memory\_HeapMemoryUsed) peaked at 1017709120.0}, which is a very high value, suggesting memory issues.}
  \trans{Answer}{\bad{1. Tomcat01}: High JVM memory usage and CPU idle spikes indicate memory exhaustion [\ldots] 2. Mysql02. 3. ServiceTest11.}
  }}%
  }
    \caption{\textbf{Two failed trajectories of Qwen3-4B.} Blue marks evidence or intermediate conclusions that the final answer does not follow, and red marks the final answer and its evidence.}
  \label{fig:behavior}
\end{figure}

\begin{itemize}[leftmargin=*]
  \item \textbf{Behavior-1: Reliance on large anomaly scores.} In 487 of the 640 trajectories (76\,\%), the model examined multiple candidates but gave a large anomaly score as the rationale for its final selection. In Fig.~\ref{fig:behavior-a}, after checking Tomcat, IG, and MG, it selects Tomcat01 for its high JVM memory usage and repeated GC logs.
  \item \textbf{Behavior-2: Inconsistency with propagation analysis.} In 433 trajectories (68\,\%), the final answer departed from the conclusion of the model's own propagation analysis~\cite{stalled-forge26}. In Fig.~\ref{fig:behavior-a}, the model suspects a problem upstream of the Tomcats but still selects the downstream Tomcat01.
  \item \textbf{Behavior-3: Misattributed signals.} In 422 trajectories (66\,\%), the model attributed a signal of another component to its answer. In Fig.~\ref{fig:behavior-b}, it cites the JVM heap signal of IG01 as ``Tomcat01's JVM memory usage'' and selects Tomcat01.
\end{itemize}

\begin{autowrap}{table}{0.45\columnwidth}
  \centering
  \caption{\textbf{Accuracy with a few-shot procedure.} Name: answers matching the example's root cause. Family: same, ignoring instance numbers.}
  \label{tab:masking}

  \footnotesize
  \setlength{\tabcolsep}{1.5pt}
  \begin{tabular}{
    @{}
    >{\raggedright\arraybackslash}p{0.51\columnwidth}
    c@{\hspace{6pt}}c@{\hspace{6pt}}c
    @{}
  }
    \toprule
    \textbf{Condition} & \textbf{Top-1} & \textbf{Name} & \textbf{Family} \\
    \midrule
    No injection
      & 16\,\% & -- & -- \\
    Same cause, names kept
      & 56\,\% & 56\,\% & 81\,\% \\
    Same cause, names masked
      & 21\,\% & -- & -- \\
    Different cause, names kept
      & 9\,\% & 44\,\% & 67\,\% \\
    \bottomrule
  \end{tabular}
\end{autowrap}

\noindent\textbf{Effect of procedure injection.}
To test whether providing a procedure changes this selection, we appended to $p$ a few-shot example~\cite{brown2020languagemodelsfewshotlearners} in which DeepSeek-R1~\cite{deepseek} reasons step by step to the correct answer of another incident. When the example shares the root-cause component of the target, Top-1 (correct first answers) rises from 16\,\% to 56\,\%, but masking the component names in the example lowers it to 21\,\% (Table~\ref{tab:masking}). When the causes differ, 44\,\% of the answers copy the root-cause name of the example, and Top-1 falls to 9\,\%. The gain thus comes largely from the names rather than from the procedure. FoundRoot-14B~\cite{icse26-foundroot}, which learns the diagnostic procedure through SFT and RL, also relies on large anomaly scores (Behavior-1). On the inputs released by the authors, 72\,\% of its top-ranked answers are the component with the most anomalous metrics, and it is correct for 83\,\% of incidents when that component is the root cause but for none otherwise.

\begin{finding}{2} 
The SLM examines candidates and analyzes propagation in every trajectory, but its final answer often does not follow this analysis. Providing the diagnostic procedure as a few-shot example or through training does not remove its reliance on large anomaly scores. The model should therefore be trained to select the root cause consistent with the evidence in the prompt while keeping its existing reasoning.
\end{finding}

\section{Design}\label{sec:design}
We present an SLM-based RCA system that addresses Findings 1 and 2 by changing what the prompt states about each signal and how the SLM selects the root cause. \textbf{Historical Frequency Profiling} adds to the prompt how often each signal exceeds $\tau$ in the fault-free period, which the model cannot infer from the incident window alone (Finding 1). \textbf{Biased-Decision Correction} keeps the diagnostic procedure that the SLM already performs and trains only the decision part in order to let the SLM's own reasoning lead to correct answers (Finding 2).

\begin{figure*}[t]
  \centering
  \includegraphics[width=\textwidth]{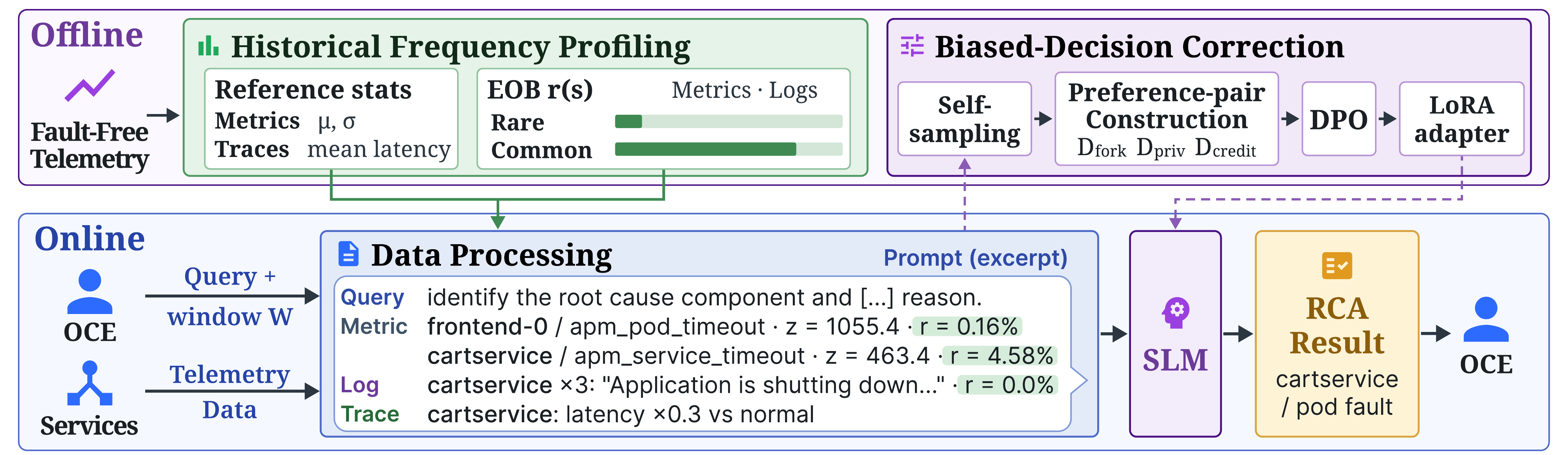}
  \caption{\textbf{System overview.} Top: offline profiling and training. Bottom: online diagnosis of one incident.}
  \label{fig:overview}
\end{figure*}

Fig.~\ref{fig:overview} shows the architecture of our system. The offline path (top) runs once per target system. It profiles the fault-free period and trains a Low-Rank Adaptation (LoRA)~\cite{lora} adapter with DPO on preference pairs built from the SLM's own trajectories. The online path (bottom) builds the prompt $p$ for an incident window $W$, attaching the EOB to each signal and ordering candidates by it, and the SLM with the adapter reads $p$ once to answer the component, fault type, and occurrence time.

\subsection{Historical Frequency Profiling}\label{sec:design-input}
Historical Frequency Profiling records, for each signal, how often it exceeds $\tau$ when no fault is present. Our system writes this frequency, the \emph{Expected Outlier Baseline} (EOB), next to the anomaly score of each signal in the prompt and orders candidates by it.

\subsubsection{Expected Outlier Baseline (EOB)}\label{sec:design-input-profile}

The EOB $r(s)$ of signal $s$ is the fraction of fault-free windows $W' \in \mathcal{W}_{\mathrm{normal}}$, defined as in \S\ref{sec:study-order}, in which its anomaly score exceeds $\tau$:
\begin{equation}
 r(s) = \Pr_{W' \in \mathcal{W}_{\mathrm{normal}}}\bigl[\, \phi(s; W') > \tau \,\bigr],
 \label{eq:eob}
\end{equation}
where $\phi(s; W')$ is computed within $W'$. Because $r(s)$ is counted directly without assuming a distribution, it places KPIs with different $\sigma$ and logs counted as entries on a single scale. The fault-free windows come from a profiling period long enough for $\mu$, $\sigma$, and the EOB to become stable, over which the normal behavior of the system is assumed to stay similar. 

For metrics, we use $\tau = 3$~\cite{ase-opsagent, fse26-rethinking} as in existing data processing but estimate $\mu$ and $\sigma$ from all of $\mathcal{W}_{\mathrm{normal}}$ rather than from the few minutes before the incident window. For logs, $\tau = 0$, so $r(s)$ is the fraction of fault-free windows in which the template of $s$ appears. Traces have no EOB. Instead, the average latency of each service over $\mathcal{W}_{\mathrm{normal}}$ is stored as a baseline, and the prompt reports the ratio of the latency in the incident window to this baseline. Profiling needs the times of past incidents to identify the fault-free period but not their root causes. At diagnosis time, our system computes $\phi(s)$ in $W$ with the profiled $\mu$ and $\sigma$ and looks up $r(s)$ for each signal that exceeds $\tau$.

\subsubsection{Candidate Ordering and Prompt Rendering}\label{sec:design-input-ordering}
At diagnosis time, the EOB serves two roles in the prompt. It is written next to the anomaly score of each signal, and it determines the order of the candidates. Prior work lists candidates without a defined order or by deviation magnitude~\cite{ase-opsagent, icse26-foundroot}, which the anomaly score already conveys. Our system instead lists first the candidates whose anomalies are rare in the fault-free period.

The prompt lists the following fields for each telemetry type, with the EOB written as a percentage (e.g., ``fires in 0.39\,\% of normal windows'').

\begin{itemize}[leftmargin=*]
  \item \textit{Metric:} KPI name, anomaly score $\phi$ (written as $z_{\max}$), EOB, and the time the KPI first exceeded $\tau$.
  \item \textit{Log:} occurrence count and EOB of each template, per component.
  \item \textit{Trace:} span count, latency ratio to normal, and the first time the ratio exceeds $\eta$, per service.
\end{itemize}
We set $\eta = 3$, the same value as $\tau$ for metrics, so that a trace is flagged only when its latency deviates markedly from normal operation. Blocks for unavailable telemetry types are omitted.

\begin{autowrap}{algorithm}{0.45\linewidth}
\small
\captionsetup{font=small}
\caption{EOB-Guided Input Construction}
\label{alg:input-construction}
\algrenewcommand\algorithmicindent{1em}
\begin{algorithmic}[1]
\Require Query, $W$, $\mathcal{T}_W$, $G$, $\mathcal{F}$
\Require Fault-free telemetry profiles
\Ensure Model input $p$
\State $M \gets$ metric observations with $\phi(s) > \tau$
\State $\mathcal{E}_W \gets$ components in $M$
\For{$e \in \mathcal{E}_W$}
    \State $M_e \gets$ observations of $e$ in $M$
    \State $r(e) \gets \min_{s \in M_e} r(s)$
    \State $\phi_{\max}(e) \gets \max_{s \in M_e} \phi(s)$
    \State Sort $M_e$ by EOB $\uparrow$
\EndFor
\State Sort $\mathcal{E}_W$ by $r(e)\uparrow$, then $\phi_{\max}(e)\downarrow$
\State $L \gets$ logs sorted by EOB $\uparrow$
\State $T \gets$ traces sorted by latency ratio $\downarrow$
\State $p \gets \mathrm{render}(W, [(e,M_e)]_{e\in\mathcal{E}_W}, G)$
\State Append $L$ and $T$ to $p$
\State Append $\mathcal{F}$ to $p$ if query requests fault type
\State Append requested JSON format to $p$
\State \Return $p$
\end{algorithmic}
\end{autowrap}

Algorithm~\ref{alg:input-construction} shows how the prompt $p$ is constructed. Lines 1--2 obtain the root-cause candidates $\mathcal{E}_W$ from the metric signals flagged as anomalous with fault-free statistics. Logs and traces do not add candidates. Lines 3--8 compute, for each component $e$, $r(e)$ as the minimum EOB of its metric signals and $\phi_{\max}(e)$ as the maximum anomaly score, and sort its signals by EOB. Line 9 then sorts the candidates in ascending order of $r(e)$, breaking ties in descending order of $\phi_{\max}(e)$. Because $r(e)$ is the EOB of the rarest signal of a component, a component whose anomalous signals all exceed $\tau$ often in the fault-free period moves down even when its anomaly scores are large. Lines 10--11 sort logs by EOB and traces by latency ratio. Lines 12--15 write the sorted candidates and observations with $\mathrm{render}$ and append the log and trace blocks, $\mathcal{F}$ when the query asks for the fault type, and the JSON output format. The SLM then reads $p$ once and selects the root cause from all three types of observations.
\wrapstuffclear

\subsection{Biased-Decision Correction}\label{sec:design-dpo} 
Even with the EOB, the SLM often bases its final answer on a large anomaly score rather than on its own analysis (Finding 2). Biased-Decision Correction samples the SLM's own reasoning, locates the points where it diverges into correct and incorrect answers, builds and verifies preference pairs from these points, and trains the decision that follows with DPO.

\subsubsection{Preference-Pair Construction}\label{sec:design-pairs}
\begin{figure*}[t]
  \centering
  \includegraphics[width=\textwidth]{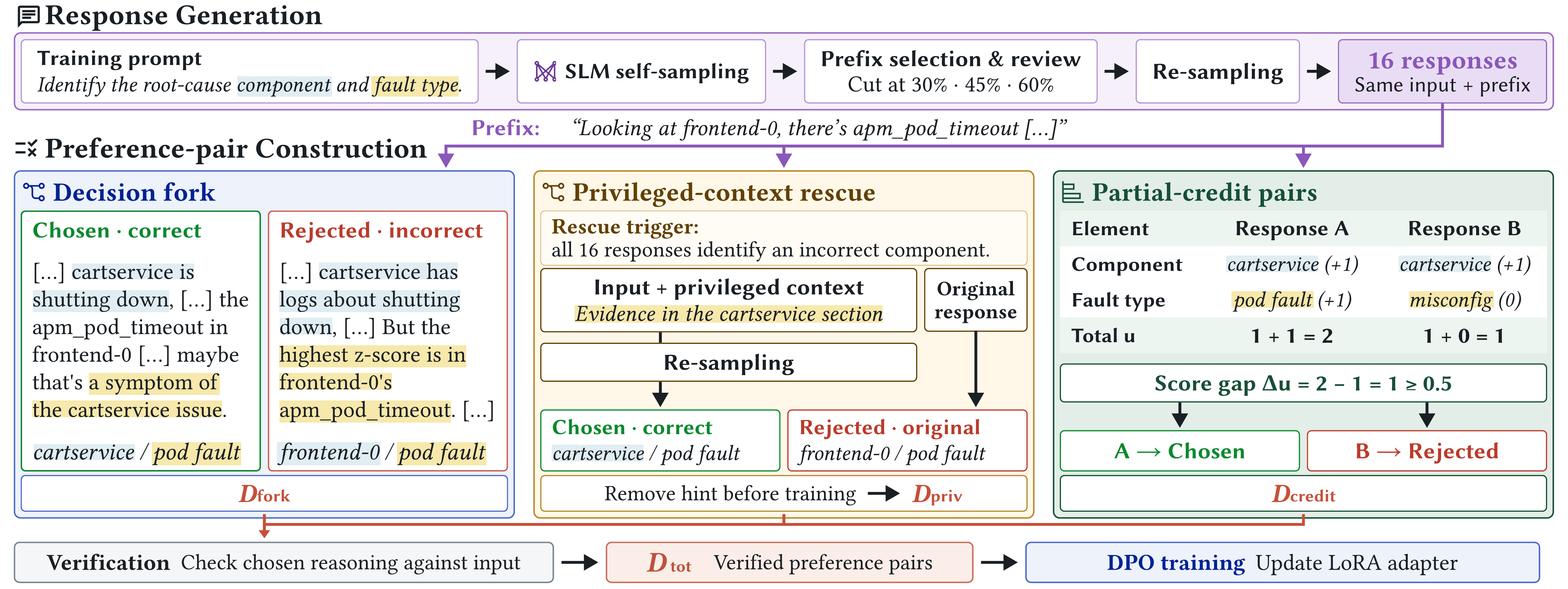}
  \caption{\textbf{Preference-pair construction for Biased-Decision Correction.} The chosen response of each pair is verified against the original input before DPO training.}
  \label{fig:pairs}
\end{figure*}

We call the model before training the reference policy $\pi_{\mathrm{ref}}$. All responses in the training pairs are generated by $\pi_{\mathrm{ref}}$, because responses from the model itself are more effective for preference training than responses corrected by an external model~\cite{arxiv24-stepdpo}. A preference pair $(x, y_w, y_l)$ consists of a shared context $x$ and a chosen and a rejected continuation, $y_w$ and $y_l$. Fig.~\ref{fig:pairs} illustrates the procedure, from self-sampling on the left to the three types of preference pairs and their verification on the right.

\textit{\textbf{Self-sampling and re-sampling.}} For each training incident, $\pi_{\mathrm{ref}}$ repeatedly generates reasoning and a final answer from its prompt $p$, forming trajectories. Datasets with fewer training incidents receive more trajectories per incident to raise the chance of observing different decisions. Among the component, fault type, and occurrence time asked by the query, one target element whose selection is to be corrected is chosen. The reasoning of each trajectory is cut at sentence boundaries at 30\,\%, 45\,\%, and 60\,\% of its length, so that prefixes end both during candidate examination and during answer composition. A reviewer LLM examines each prefix twice independently and discards those that have already fixed the answer for the target element or contain repetition, content absent from the input, or incomplete sentences. From each remaining prefix $z_{<\rho}$ of a trajectory $z$ cut at ratio $\rho$, $\pi_{\mathrm{ref}}$ generates $J$ continuations ($J = 16$ by default):
\begin{equation}
  x = p \oplus z_{<\rho}, \qquad
  y^{(1)}, \ldots, y^{(J)} \sim \pi_{\mathrm{ref}}(\cdot \mid x).
  \label{eq:fork}
\end{equation}

\textit{\textbf{Decision fork ($\mathcal{D}_{\mathrm{fork}}$).}}
When the continuations of the same context $x$ include both correct and incorrect answers for the target element, we call $x$ a \emph{decision fork}. Each pair of a correct and an incorrect continuation from $x$ is added to $\mathcal{D}_{\mathrm{fork}}$. In the decision fork example of Fig.~\ref{fig:pairs}, the target element is the component, and both continuations examine the termination log of \texttt{cartservice}. The rejected continuation, however, selects \texttt{frontend-0} based on its high $z_{\max}$, whereas the chosen continuation links the termination log to a pod fault of \texttt{cartservice} and interprets the timeout of \texttt{frontend-0} as its effect. This pair, which selects different root causes from the same prefix, is added to $\mathcal{D}_{\mathrm{fork}}$.

\textit{\textbf{Privileged-context rescue ($\mathcal{D}_{\mathrm{priv}}$).}}
When no continuation answers the target element correctly even after re-sampling, a pair cannot be formed. We then append \emph{privileged context}, auxiliary information about the correct answer, to the prompt and re-sample from the same prefix. We first give only the component that holds the evidence, or an approximate time of the fault for queries that ask for the time, and add the correct label only if this still yields no correct continuation. The resulting correct continuation becomes the chosen response and is paired with an original incorrect continuation on the original context $x$. In the rescue example of Fig.~\ref{fig:pairs}, \texttt{cartservice / pod fault} obtained this way is paired with the original \texttt{frontend-0 / pod fault}. The privileged context thus serves only to obtain the chosen response and is absent during training. Unlike STaR~\cite{neurips22-star}, which uses such rationalized solutions for SFT, we pair them with the original incorrect answers, discard continuations that mention the privileged context, and check the rest for grounding in the original input.

\textit{\textbf{Partial-credit pairs ($\mathcal{D}_{\mathrm{credit}}$).}}
Continuations with the same correctness for the target element can still differ in the other requested elements. We therefore score each continuation by $u$, the sum over the requested elements of 1 point for a correct component, 1 point for a correct fault type, and 1 point for an occurrence time within $\pm 60$ seconds (0.5 points within $\pm 300$ seconds). We pair the top two continuations of each prefix by $u$ with the bottom two and keep the pairs with $\Delta u \ge 0.5$ whose continuations agree on the correctness of the target element, because the other pairs are already covered by $\mathcal{D}_{\mathrm{fork}}$. In the partial-credit example of Fig.~\ref{fig:pairs}, both continuations identify \texttt{cartservice}, but only the one that answers \texttt{pod fault} identifies the fault type, so the response with $u=2$ is chosen over the one with $u=1$. The score $u$ only selects pairs and does not weight the loss.

\textit{\textbf{Verification.}}
A correct final answer may come with an explanation that is not grounded in the input. Therefore, the chosen responses of the three sets are reviewed twice independently by an LLM. Each review checks whether a response misattributes the observations of another component, presents values or times absent from the input, or misreads a metric's value or meaning. If either of the two judgments reports a problem, the preference pair is removed.

\subsubsection{DPO Training}\label{sec:design-dpo-obj}

Biased-Decision Correction trains the target model $\pi_\theta$ with DPO~\cite{neurips23-dpo} on the preference pairs that pass verification, $\mathcal{D}_{\mathrm{total}} = \mathcal{D}_{\mathrm{fork}} \cup \mathcal{D}_{\mathrm{priv}} \cup \mathcal{D}_{\mathrm{credit}}$. Without a separate reward model, DPO minimizes the following loss so that the chosen response is preferred over the rejected response in the same context:
\begin{equation}
  \mathcal{L}_{\mathrm{DPO}}
  = -\mathbb{E}_{(x,y_w,y_l)\sim\mathcal{D}_{\mathrm{total}}}
  \Bigl[\log\operatorname{sigmoid}\Bigl(
    \beta\log\frac{\pi_\theta(y_w\mid x)}{\pi_{\mathrm{ref}}(y_w\mid x)}
    -\beta\log\frac{\pi_\theta(y_l\mid x)}{\pi_{\mathrm{ref}}(y_l\mid x)}
  \Bigr)\Bigr],
  \label{eq:dpo}
\end{equation}
where $\beta$ controls the deviation from $\pi_{\mathrm{ref}}$. Because the input and the prefix serve only as the condition $x$, the loss is computed only on the tokens of the continuations. The training signal thus compares which reasoning and answer should follow the prefix rather than teaching the model to reproduce the prefix. $\mathcal{D}_{\mathrm{total}}$ is constructed before training, no new responses are generated during training, and every pair has the same weight.
 
Training uses LoRA~\cite{lora}, which freezes the weights of the base model and updates only the adapter. We train one adapter per dataset on all projections of the attention and MLP layers (rank 16, $\alpha=32$, dropout 0.05), with $\beta=0.1$, a learning rate of $5\times10^{-6}$ under a cosine schedule with 10\,\% warmup, 4 epochs, and an effective batch size of 8. The reference policy is the base model without the adapter, and the prompts for both training and evaluation are constructed by Algorithm~\ref{alg:input-construction}.

\section{Evaluation}\label{sec:eval}
We evaluate our system by answering the following five research questions.

\begin{itemize}[leftmargin=*]
  \item \textbf{RQ1 (Effectiveness):} How accurately does our system perform RCA compared with existing RCA systems that use commercial or fine-tuned models? (\S\ref{sec:eval-main})
  \item \textbf{RQ2 (Ablation Study):} How much do the EOB and the three types of preference pairs each contribute to accuracy? (\S\ref{sec:eval-ablation})
  \item \textbf{RQ3 (Generalizability):} Does an adapter trained on one dataset generalize, without further fine-tuning, to held-out datasets and systems? (\S\ref{sec:eval-foundroot})
  \item \textbf{RQ4 (Model, Model Size, and Quantization):} How do the base model, its size, and quantization affect the accuracy and deployment efficiency of our system? (\S\ref{sec:eval-size})
  \item \textbf{RQ5 (Deployment Overhead):} How much fault-free history and offline computation does deploying our system on a target system require? (\S\ref{sec:eval-sensitivity})  
\end{itemize}

\subsection{Experimental Setup}\label{sec:eval-setup}

\subsubsection{Datasets}

\begin{autowrap}{table}{0.42\columnwidth}
  \centering
  \caption{\textbf{Datasets and their training and evaluation incidents.}}
  \label{tab:datasets}

  \footnotesize
  \setlength{\tabcolsep}{1.5pt}

  \begin{tabularx}{\linewidth}{
    @{}
    l
    >{\raggedright\arraybackslash}X
    r@{\hspace{1em}}r
    @{}
  }
    \toprule
    \textbf{Dataset} & \textbf{System} & \textbf{Train} & \textbf{Eval.} \\
    \midrule

    \multicolumn{4}{@{}l}{\textit{\textbf{Training and evaluation within each system}}} \\
    Telecom~\cite{openrca}   & Telecom service   & 30  & 21  \\
    Bank~\cite{openrca}      & Banking service   & 80  & 56  \\
    Market~\cite{openrca}    & HipsterShop & 88 & 60  \\
    AIOps 2025~\cite{aiops2025} & HipsterShop    & 241 & 159 \\

    \midrule

    \multicolumn{4}{@{}l}{\textit{\textbf{Evaluation on held-out datasets}}} \\
    DejaVu-A~\cite{fse22-dejavu}
      & Major ISP system & -- & 63 \\
    DejaVu-B~\cite{fse22-dejavu}
      & Major ISP system & -- & 90 \\
    Eadro-SN~\cite{icse23-eadro}
      & SocialNetwork & -- & 10 \\
    GAIA~\cite{GAIA}
      & MicroSS & -- & 255 \\

    \bottomrule
  \end{tabularx}

  \par\smallskip
  {\raggedright\scriptsize A dash marks datasets not used for training.\par}
\end{autowrap}

Table~\ref{tab:datasets} summarizes the datasets and their training and evaluation incidents. Telecom, Bank, and Market of OpenRCA~\cite{openrca} are collected from a telecommunication database system, a banking system, and the e-commerce application HipsterShop, respectively, and AIOps 2025~\cite{aiops2025} is collected by injecting faults into HipsterShop deployed on Kubernetes. On these four datasets, we train an adapter on the training incidents of each system and measure accuracy on its evaluation incidents by randomly splitting the incidents of each dataset into 60\,\% for training and 40\,\% for evaluation.

We evaluate generalization across systems on DejaVu-A and DejaVu-B~\cite{fse22-dejavu}, Eadro-SN~\cite{icse23-eadro}, and GAIA~\cite{GAIA}, the four held-out test datasets of FoundRoot, none of which FoundRoot or our system uses for training. The GAIA test set released by FoundRoot labels only the root-cause component, so only the component is evaluated on it.

\subsubsection{Evaluation Metrics}\label{sec:eval-metrics}
Following OpenRCA~\cite{openrca}, we score only the component, fault type, and occurrence time that the query requests. The component is compared at the service or instance level specified by the label, and the fault type must match the label in the list of fault types of the benchmark. The occurrence time is correct if it is within 60 seconds of the label. For queries that ask about multiple faults, predicted faults are matched to labeled faults before scoring.

\textbf{\Exact} is the ratio of responses that correctly answer all elements the query requests, which OpenRCA calls Correct~\cite{openrca}. \textbf{\Partial} is the average fraction of requested elements answered correctly per response. This definition differs from the Partial of OpsAgent~\cite{ase-opsagent}, which counts a response as partially correct if it answers at least one requested element. \textbf{Component}, \textbf{Fault Type}, and \textbf{Time} score each element separately. Among the queries that request the element, each is the fraction of responses that answer it correctly, where a response to a multi-fault query counts as correct if it matches at least one labeled value.

\subsubsection{Baselines}
We compare our system with three groups of baselines.

\begin{itemize}[leftmargin=*]
  \item \textbf{Statistical RCA.} Nezha~\cite{fse23-nezha} localizes the root cause from changes in event patterns constructed from metrics, logs, and traces. KPIRoot~\cite{issre24-kpiroot} ranks KPIs from metric time series by combining the symbolic sequence similarity between the alert KPI and candidate KPIs with Granger causality. KPIRoot+~\cite{kpiroot+} extends KPIRoot by detecting anomalous intervals with STL decomposition and a U-Net autoencoder representing change trends with iSAX~\cite{isax}, a symbolic time-series encoding.
  \item \textbf{LLM-based RCA without fine-tuning.} CoT~\cite{cot} prompts the off-the-shelf Qwen3-4B to reason step by step over the prompt built by the data processing of OpsAgent~\cite{ase-opsagent}. ReAct~\cite{react} lets the same model alternate, for up to 10 turns, between reasoning and calls to the trace, log, and metric query tools released with ThinkFL~\cite{tosem-thinkfl}. RCA-Agent~\cite{openrca}, designed for commercial models, investigates telemetry with a Controller that plans the analysis and an Executor that runs generated Python code. We run it with GPT-5-mini~\cite{gpt-5} and Solar Pro 4~\cite{solar-pro-4}.
  \item \textbf{RCA with fine-tuned models.} OpsAgent-14B~\cite{ase-opsagent} trains expert agents with PPO and has them cross-review their diagnoses. FoundRoot-14B~\cite{icse26-foundroot} trains structured deep thinking with SFT and RL to rank components from metrics. ThinkFL-8B~\cite{tosem-thinkfl} fine-tunes Llama with SFT on tool-calling conversations and then with GRPO to query traces and metrics from the entry span of a trace.
\end{itemize}

We run Nezha with the authors' implementation after reproducing its reported performance, and KPIRoot with its public code with minor fixes. We implement KPIRoot+ by adding the three changes described in its paper to this code, and both use the service-level objective metric of each system as the alert KPI~\cite{issre24-kpiroot, kpiroot+}. For RCA-Agent, we score one run of the public code per incident~\cite{kim2026aiagentssystematicallyfail}, and for OpsAgent, the authors' released results on our evaluation incidents. FoundRoot and ThinkFL use their released checkpoints (and tools for ThinkFL), with fault type and occurrence time fields added to their prompts and answer formats because both output only components.

\subsubsection{Implementation}
The default backbone of our system is Qwen3-4B~\cite{qwen3} in thinking mode. Historical Frequency Profiling uses the entire fault-free period of each dataset, and \S\ref{sec:eval-sensitivity} evaluates shorter profiling periods. For self-sampling, the reference policy generates trajectories at temperature 0.6, 32 per training incident for Telecom, Bank, and Market and 8 per training incident for AIOps 2025. Claude Opus 5~\cite{claude} serves as the reviewer LLM for prefix review and verification. The reviewer is used only offline to build training pairs from benchmark data, and online diagnosis sends no telemetry outside the organization. In production, a self-hosted model can take this role.

Biased-Decision Correction trains the LoRA adapter~\cite{lora} with the DPOTrainer of TRL~\cite{hf-dpotrainer} under the settings in \S\ref{sec:design-dpo-obj}, with a maximum sequence length of 12,288 tokens (18,432 for Market) so that no pair is truncated. Qwen3-4B is trained on a single NVIDIA L40S (48\,GB), and the 14B model in RQ4 on a single NVIDIA A100 (80\,GB). Inference runs on vLLM~\cite{vllm} with temperature 0.6, top-p 0.95, top-k 20, and a maximum generation length of 6,144 tokens. We generate eight responses per incident, score each independently, and report their average (avg@8)~\cite{ase-opsagent, fse26-torai, fse24-baro, Kimi-K2}. \emph{Off-the-shelf} denotes the same base model with the same generation settings but without the LoRA adapter. Unless noted otherwise, it receives the EOB prompt.

\subsection{RQ1: Effectiveness}\label{sec:eval-main}

\newlength{\gseplen}\setlength{\gseplen}{8pt}
\newcommand{\gsep}{\hspace{\gseplen}}
\newlength{\rpad}\setlength{\rpad}{6pt}
\newlength{\gshift}\setlength{\gshift}{1.5pt}
\newlength{\grulew}\newlength{\grulet}

\newcommand{\ghead}[1]{\multicolumn{2}{c@{\gsep}}{\makebox[0pt]{\textbf{#1}}\hspace*{2\gshift}}}
\newcommand{\gheadlast}[1]{\multicolumn{2}{c@{\hspace{\rpad}}}{\makebox[0pt]{\textbf{#1}}\hspace*{2\gshift}}}

\newcommand{\grouprule}[1]{%
  \addlinespace[2.5pt]%
  \multicolumn{9}{@{}l@{}}{\textit{#1}\hspace{.6em}\xrfill[.45ex]{\lightrulewidth}}\\[-.3ex]}
\newcommand{\grouprulefirst}[1]{%
  \multicolumn{9}{@{}l@{}}{\textit{#1}\hspace{.6em}\xrfill[.45ex]{\lightrulewidth}}\\[-.3ex]}

\begin{table*}[t]
\centering
\caption{\textbf{Accuracy on four datasets (\%).}
\Exact: ratio of responses that answer all requested elements correctly. \Partial: average fraction of requested elements answered correctly per response.}
\label{tab:baselines}
\footnotesize
\setlength{\tabcolsep}{3pt}
\renewcommand{\arraystretch}{0.95}
\settowidth{\grulew}{\textbf{AIOps2025 (159)}}%
\settowidth{\grulet}{\textbf{\Exact}\textbf{\Partial}}%
\setlength{\grulet}{\dimexpr(\grulet+2\tabcolsep-\grulew)/2-1.5pt\relax}%
\begin{tabularx}{\linewidth}{@{}X cc @{\gsep} cc @{\gsep} cc @{\gsep} cc @{\hspace{\rpad}}}
\toprule
\textbf{Method} & \ghead{Telecom} & \ghead{Bank} & \ghead{Market} & \gheadlast{AIOps 2025} \\
\cmidrule(l{\dimexpr3pt+\grulet-\gshift\relax}r{\dimexpr\gseplen+\grulet+\gshift\relax}){2-3}
\cmidrule(l{\dimexpr3pt+\grulet-\gshift\relax}r{\dimexpr\gseplen+\grulet+\gshift\relax}){4-5}
\cmidrule(l{\dimexpr3pt+\grulet-\gshift\relax}r{\dimexpr\gseplen+\grulet+\gshift\relax}){6-7}
\cmidrule(l{\dimexpr3pt+\grulet-\gshift\relax}r{\dimexpr\rpad+\grulet+\gshift\relax}){8-9}
 & \textbf{\Exact} & \textbf{\Partial} & \textbf{\Exact} & \textbf{\Partial} & \textbf{\Exact} & \textbf{\Partial} & \textbf{\Exact} & \textbf{\Partial} \\[-.4ex]
\grouprulefirst{\textbf{Statistical RCA}}
Nezha~\cite{fse23-nezha}$^{\dagger}$ & 0.0 & 2.4 & 1.8 & 4.8 & 0.0 & 1.2 & 0.0 & 4.4 \\
KPIRoot~\cite{issre24-kpiroot} & 0.0 & 2.4 & 1.8 & 1.8 & 1.7 & 3.6 & 0.0 & 2.5 \\
KPIRoot+~\cite{kpiroot+} & 0.0 & 6.3 & 0.0 & 0.9 & 1.7 & 2.5 & 0.0 & 4.2 \\
Statistical EOB (Ours) & 0.0 & 10.3 & 7.1 & 17.7 & 0.0 & 0.3 & 0.0 & 15.3 \\
\grouprule{\textbf{LLM-based RCA without fine-tuning}}
CoT (Qwen3-4B)~\cite{cot} & 6.5 & 15.0 & 2.5 & 7.9 & 3.8 & 12.9 & 0.5 & 20.3 \\
ReAct (Qwen3-4B)~\cite{react} & 3.0 & 10.8 & 9.6 & 15.2 & 4.2 & 13.6 & 2.3 & 19.1 \\
RCA-Agent~\cite{openrca} (GPT-5-mini~\cite{gpt-5})$^{\ddagger}$ & 14.3 & 23.8 & 7.1 & 12.9 & 5.0 & 17.4 & 2.5 & 22.3 \\
RCA-Agent~\cite{openrca} (Solar Pro 4~\cite{solar-pro-4}) & 4.8 & 15.9 & 12.5 & 21.4 & 1.7 & 9.7 & 8.2 & 26.1 \\
\grouprule{\textbf{RCA with fine-tuned models}}
OpsAgent-14B~\cite{ase-opsagent}$^{\S}$ & 30.0 & 37.5 & 18.5 & 27.6 & \textbf{11.9} & 23.0 & N/A & N/A \\
ThinkFL-8B~\cite{tosem-thinkfl}$^{\|}$ & 0.0 & 0.0 & 0.0 & 0.0 & 0.0 & 1.8 & 0.0 & 8.0 \\
\textbf{Ours-4B} & \textbf{39.3} & \textbf{55.3} & \textbf{48.7} & \textbf{55.5} & \textbf{11.9} & \textbf{24.3} & \textbf{21.1} & \textbf{52.4} \\
\bottomrule
\end{tabularx}
\par\smallskip
\parbox{\linewidth}{\scriptsize\raggedright
\mbox{$^{\dagger}$Outputs not mapped to benchmark fault types.} \quad
\mbox{$^{\ddagger}$Public code, one run per incident~\cite{kim2026aiagentssystematicallyfail}.} \quad
\mbox{$^{\S}$Authors' result files (Qwen2.5-14B); no AIOps 2025.} \quad
\mbox{$^{\|}$Llama 3 8B trained on HipsterShop (Market, AIOps 2025).}}
\end{table*}

Against all baselines on the four datasets (Table~\ref{tab:baselines}), our system achieves the highest \Partial everywhere and the highest \Exact on all but Market, where it ties with OpsAgent-14B.

The statistical methods do not output the fault types or occurrence times of the benchmarks, so they cannot complete such queries, and the \Exact of Nezha and the KPIRoot family stays below 2\,\% on all four datasets. Statistical EOB, which returns the first candidate in the EOB-ordered list, achieves higher \Partial than the other statistical methods on Telecom, Bank, and AIOps 2025, but its \Exact is nonzero only on Bank, where some queries request only the component. The gap from our system, which reads the same prompt to identify all requested elements, shows the role of the model in completing the diagnosis from this evidence.

Compared with LLM-based methods without fine-tuning, our system achieves higher \Exact and \Partial on all datasets than CoT and ReAct, which use the same Qwen3-4B, and than RCA-Agent with GPT-5-mini and Solar Pro 4. Market is the most challenging dataset, where no method exceeds 11.9\,\% \Exact. The difficulty concentrates on two-fault queries, which make up 45\,\% of Market (27 of 60) but at most 11\,\% of the other datasets. Every baseline answers at most 4.0\,\% of these queries correctly, whereas our system answers 9.7\,\%. On single-fault queries, OpsAgent-14B leads instead (17.6\,\% vs.\ 13.6\,\%), so the two methods tie on Market overall.

OpsAgent-14B, in which expert agents trained with PPO diagnose and review each other's results~\cite{ase-opsagent}, is the strongest baseline on Telecom, Bank, and Market. Our system outperforms it on both metrics on Telecom and Bank and achieves a higher \Partial on Market. ThinkFL is trained on HipsterShop~\cite{tosem-thinkfl}, the system of Market and AIOps 2025, yet its \Exact is 0 on both. Its \Partial comes only from the ranked components, and none of its responses gives the fault type or occurrence time, even though its prompt and answer format request them.

\begin{answer}{1}
With a 4B model and without cross-agent review or repeated tool calls on raw telemetry, our system matches or exceeds every baseline on both metrics across all four datasets, including the 14B OpsAgent and RCA-Agent with commercial models.
\end{answer}

\subsection{RQ2: Ablation Study}\label{sec:eval-ablation}

\begin{autowrap}{figure}{0.46\linewidth}
  \centering
  \includegraphics[width=\linewidth]{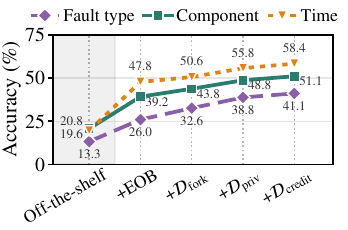}
  \caption{\textbf{Per-element accuracy as components are added cumulatively (avg@8, \%).}}
  \label{fig:ablation}
\end{autowrap}

\noindent To see how each component of our system contributes, Fig.~\ref{fig:ablation} adds the components of our system cumulatively from left to right on the x-axis and shows the Component, Fault Type, and Time scores on the four evaluation splits. The leftmost step is the off-the-shelf Qwen3-4B on the prompt of existing data processing. +EOB replaces this prompt with the EOB prompt, and each later step adds one pair type and trains a new adapter from the base model on all pairs added so far.

\noindent\textbf{+EOB.} Without fine-tuning, the EOB prompt raises all three elements. Component rises because the root cause is listed first more often. On Bank, the first candidate in the EOB prompt is the root cause for two thirds of incidents, whereas the component with the largest anomaly score in the existing prompt is the root cause for only 8\,\%. Time rises the most because the EOB prompt states, for each candidate, the first time its rarest signal exceeds $\tau$, whereas answering the first timestamp in the existing prompt is correct for fewer than 20\,\% of incidents. Fault Type stays the lowest of the three because the prompt does not point to it, and it requires diagnosing what the signals of the chosen component indicate.

\noindent\textbf{\boldmath +$\mathcal{D}_{\mathrm{fork}}$.} These pairs make the model reach, in more of its responses, a correct decision that it previously reached in only some of them. Because accuracy is averaged over eight responses per incident, most of the gain at this step on every dataset comes from such incidents.

\noindent\textbf{\boldmath +$\mathcal{D}_{\mathrm{priv}}$.} Contexts in which no continuation is correct give $\mathcal{D}_{\mathrm{fork}}$ no chosen response, and $\mathcal{D}_{\mathrm{priv}}$ supplies one through privileged context, which is absent from the training input. These pairs show the model a line of reasoning from the evidence in its own prompt to the full answer, and all three elements rise at this step. On Bank, for example, \Exact rises from 33.0\,\% to 47.1\,\%.

\noindent\textbf{\boldmath +$\mathcal{D}_{\mathrm{credit}}$.} $\mathcal{D}_{\mathrm{fork}}$ and $\mathcal{D}_{\mathrm{priv}}$ separate correct answers from incorrect ones but do not guide a partly correct response, such as one with the right component, toward the remaining elements. $\mathcal{D}_{\mathrm{credit}}$ fills this gap by preferring, between two responses that share part of the answer, the one that answers more requested elements correctly, and all three elements rise at this step on average, although Component on Bank and Fault Type on Market fall by 1.5 and 3.9 points.

\begin{answer}{2}
Every component raises accuracy in a different way. The EOB prompt alone ranks the root cause higher and shows when each anomaly begins. Among the pairs, $\mathcal{D}_{\mathrm{fork}}$ makes correct decisions consistent, $\mathcal{D}_{\mathrm{priv}}$ covers incidents never answered correctly, and $\mathcal{D}_{\mathrm{credit}}$ completes partly correct responses.
\end{answer}

\subsection{RQ3: Generalizability}\label{sec:eval-foundroot}

\begin{autowrap}{table}{0.5\linewidth}[hsep=1em]
\centering
\caption{\textbf{Accuracy on the held-out test datasets of FoundRoot (\%).} Each method uses its own data processing. Comp.: Component. Type: Fault Type.}
\label{tab:foundroot}
\scriptsize
\setlength{\tabcolsep}{2.5pt}
\begin{tabular}{@{}l l cc|ccc@{}}
\toprule
\textbf{Dataset} & \textbf{Method} & \textbf{\Exact} & \textbf{\Partial} & \textbf{Comp.} & \textbf{Type} & \textbf{Time} \\
\midrule
\multirow{2}{*}{\makecell[l]{DejaVu-A\\(63)}} & FoundRoot-14B & 1.6 & 22.2 & 41.3 & 12.7 & 12.7 \\
 & Ours-4B & \textbf{10.7} & \textbf{50.9} & \textbf{44.2} & \textbf{28.8} & \textbf{79.6} \\
\addlinespace
\multirow{2}{*}{\makecell[l]{DejaVu-B\\(90)}} & FoundRoot-14B & 0.0 & 21.9 & 38.9 & 12.2 & 14.4 \\
 & Ours-4B & \textbf{6.4} & \textbf{38.2} & \textbf{40.4} & \textbf{25.6} & \textbf{48.8} \\
\addlinespace
\multirow{2}{*}{\makecell[l]{Eadro-SN\\(10)}} & FoundRoot-14B & 40.0 & 66.7 & \textbf{70.0} & 70.0 & 60.0 \\
 & Ours-4B & \textbf{70.0} & \textbf{76.7} & \textbf{70.0} & \textbf{71.3} & \textbf{88.8} \\
\addlinespace
\multirow{2}{*}{\makecell[l]{GAIA\\(255)}} & FoundRoot-14B & -- & -- & \textbf{88.6} & -- & -- \\
 & Ours-4B & -- & -- & 68.6 & -- & -- \\
\bottomrule
\end{tabular}
\par\smallskip
\parbox{\linewidth}{\scriptsize\raggedright (n): number of incidents. --: GAIA has no fault type or time labels.}
\end{autowrap}

\noindent To evaluate whether our system generalizes to systems unseen in training, we apply the adapter trained on the $\mathcal{D}_{\mathrm{total}}$ pairs of Bank to the four held-out datasets without further fine-tuning and compare it with FoundRoot-14B in Table~\ref{tab:foundroot}. Our system builds its prompt with Algorithm~\ref{alg:input-construction} and computes the EOB from the fault-free period of each target system, while FoundRoot builds its input for the same incident windows with its own data processing.

On DejaVu-A, DejaVu-B, and Eadro-SN, the two methods differ more in fault type and time accuracy than in component accuracy. On Eadro-SN, both identify the root-cause component in 70.0\,\% of the incidents, but our system identifies the occurrence time more often (88.8\,\% vs.\ 60.0\,\%) and raises \Exact from 40.0\,\% to 70.0\,\%. On DejaVu-A and DejaVu-B, higher Fault Type and Time raise \Partial from 22.2\,\% to 50.9\,\% and from 21.9\,\% to 38.2\,\%, although \Exact stays at most 10.7\,\% because Component and Fault Type remain below 45\,\% and 30\,\%.

This gap in fault type and time accuracy reflects the design of our system, whose prompt states when each signal first exceeds $\tau$ and whose preference pairs reward every requested element. FoundRoot instead learns only to rank root-cause components~\cite{icse26-foundroot}, a strength that shows on GAIA, which evaluates only the component (88.6\,\% vs.\ 68.6\,\%). On the other three datasets, our system reaches component accuracy comparable to FoundRoot's with a 4B adapter trained on one system while also identifying the fault type and time.

Bank and the target systems differ in component names and fault type notation but share phenomena such as CPU load and network faults. Our system receives the candidates and fault types of each target system, and the EOB is a ratio to each system's own fault-free period, so its values mean the same across systems. The adapter can therefore reuse the judgment learned on Bank to select the root cause from observations.

\begin{answer}{3}
Trained on Bank, the 4B adapter transfers to unseen systems without further fine-tuning. On three of them, it matches FoundRoot-14B in component accuracy and exceeds it in \Exact and \Partial by identifying the fault type and time, while FoundRoot-14B leads on GAIA, which evaluates only components.
\end{answer}

\subsection{RQ4: Model, Model Size, and Quantization}\label{sec:eval-size}

\begin{autowrap}{figure}{0.49\linewidth}[hsep=1em]
  \centering
  \includegraphics[width=\linewidth]{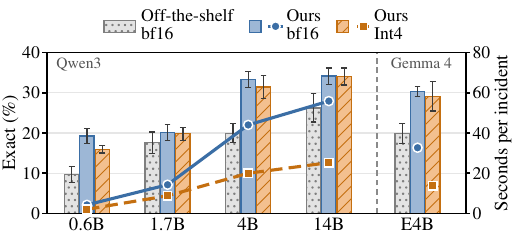}
  \caption{\textbf{\Exact (bars) and latency per incident (lines) by model and size.} All models, including Off-the-shelf, receive the EOB prompt. Error bars show the standard deviation over eight runs with one response per incident. Int4: GPTQ W4A16.}
  \label{fig:size}
\end{autowrap}

\noindent To compare deployment configurations on a single GPU, Fig.~\ref{fig:size} evaluates Off-the-shelf and Ours in bf16 and Ours in GPTQ Int4 at four Qwen3 sizes, and the same three configurations on Gemma 4 E4B~\cite{gemma4}, a model of another family. \Exact is averaged over the three OpenRCA datasets, and latency is measured on one NVIDIA L40S. All configurations use the same EOB prompt, and our system trains on preference pairs from the trajectories of each model. For Int4, we merge the LoRA adapter into the base model in float32 and apply GPTQ~\cite{gptq} W4A16 with group size 128, calibrated on training prompts.

Our system raises the mean \Exact of Qwen3-4B from 20\,\% to 33\,\%, surpassing the off-the-shelf 14B model at 26\,\% and approaching the fine-tuned 14B model at 34\,\% with 7.8 instead of 28.0\,GiB of weights. The 0.6B and 1.7B models reach only 19--20\,\% after training. Training also raises Gemma 4 E4B from 18\,\% to 30\,\%. Gemma 4 E4B answers faster than Qwen3-4B (32.7 vs.\ 44.0 seconds per incident in bf16) because it generates fewer tokens per response, but its \Exact is lower.

GPTQ Int4 reduces weight size and latency for every evaluated model. For 4B, \Exact decreases from 33\,\% to 31\,\%, while the weights shrink from 7.8 to 2.5\,GiB and latency falls from 44.0 to 20.0 seconds per incident. For 14B, Int4 keeps the bf16 \Exact with 9.3\,GiB of weights and cuts latency from 55.9 to 25.3 seconds, below that of bf16 4B, consistent with prior reports that GPTQ 4-bit can change accuracy in either direction~\cite{ijcai25-quantization-tradeoffs, neurips24-accuracy-not-all}. 14B Int4 is thus an option when accuracy matters more than the smaller weights of 4B Int4.

\begin{answer}{4}
Biased-Decision Correction improves accuracy at every evaluated model size and on a model of another family. Our system with 4B surpasses the off-the-shelf 14B model and approaches the fine-tuned 14B model with smaller weights and lower latency. Int4 further reduces weight size and latency, while preserving bf16 accuracy at 14B.
\end{answer}

\subsection{RQ5: Deployment Overhead}\label{sec:eval-sensitivity}
Deploying our approach on a new system requires fault-free history and one offline run, so we measure how much fault-free window it needs and how long training takes.

\begin{autowrap}{figure}{0.49\linewidth}
\centering
\includegraphics[width=\linewidth]{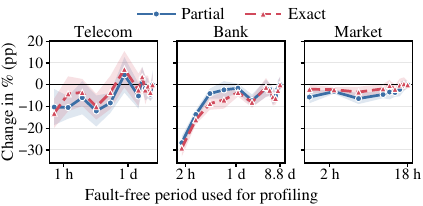}
\caption{\textbf{Accuracy change by profiling period length.} Relative to the entire fault-free period. Shading shows $\pm 1$ standard deviation over eight runs with one response per incident.}
\label{fig:sensitivity}
\end{autowrap}

Fig.~\ref{fig:sensitivity} shows how \Exact and \Partial change with a shorter fault-free period. For each incident, we keep the fault-free windows nearest the incident, before or after it, rerun Historical Frequency Profiling on these windows, and evaluate our 4B model with the same adapter. The horizontal axis is the length of the kept windows on a log scale, and its right end is the entire fault-free period, which is 80, 211, and 18 hours for Telecom, Bank, and Market. A negative value on the vertical axis means a loss.

Averaged over the three datasets, shortening the profiling period to about half a day lowers \Exact by at most 3.4 percentage points. The largest loss is 29.1 points on Bank with 1.7 hours, where only 10 fault-free windows remain per incident instead of 1,269. With $\mu$ and $\sigma$ estimated from these windows, more metrics are flagged as anomalous in $W$ (123.1 instead of 53.6 per incident), and 27.7\,\% of them receive an EOB of 0 instead of 0.5\,\%, so they appear as rare as metrics that exceed $\tau$ only during failures.

For the four datasets, the offline path of our system with Qwen3-4B takes 70.3 hours in total. Historical Frequency Profiling takes 0.4 hours on one CPU, preference-pair construction takes 13.2 hours on one L40S, and DPO training on the resulting 5,688 pairs takes 56.7 hours on one L40S. The pair-construction time is the inference time of self-sampling and re-sampling, estimated from the measured time per response.

\begin{answer}{5}
Using more of the fault-free period helps, but our system does not need the entire period. Averaged over the three datasets, \Exact decreases by at most 3.4 points when the profiling period is shortened to about half a day. Building our system with Qwen3-4B for the four datasets takes 70.3 hours, mostly for DPO training, on a single 48\,GB GPU rather than a multi-GPU cluster.
\end{answer}

\subsection{Case Study}\label{sec:eval-reasoning}
Fig.~\ref{fig:reasoning-case} compares the cause selection of the off-the-shelf Qwen3-4B and our system on the same EOB input. The incident, \texttt{qa\_229} of AIOps 2025, is a fault injected at the node level into the Kubernetes cluster that runs HipsterShop (\S\ref{sec:eval-setup}). The query gives the 30-minute window from 18:00 to 18:30 on June 17, 2025, states that one failure occurred in it, and asks for the root-cause component, the fault type, and the occurrence time, with a list of fault types that includes node fault, pod fault, and network attack. The labels are the node \texttt{aiops-k8s-03}, \texttt{node fault}, and 18:10:43. The prompt built by Algorithm~\ref{alg:input-construction} lists 28 candidates in ascending order of EOB with the fields of \S\ref{sec:design-input-ordering}. The first candidate is the pod \texttt{checkoutservice-1}: its request and response counts exceeded $\tau$ at 18:14 and do so in 0.00\,\% of the fault-free windows, and its maximum response time has an anomaly score of 362. The node is second. Its available memory (\texttt{infra\_node\_node\_memory\_MemAvailable\_bytes}) exceeded $\tau$ at 18:11 with an anomaly score of 3.5 and an EOB of 0.00\,\%, and its memory usage rate, disk read time, and disk read bytes exceeded $\tau$ in the same minute with EOBs between 0.05 and 1.01\,\%. \texttt{frontend} is the 13th candidate, with an EOB of 1.92\,\% for its rarest metric. The log block holds two entries of \texttt{frontend}, ``failed to complete the order: rpc error: code = Internal'' and the same with ``Unavailable'', each written once in the window and in 0.0\,\% of the fault-free windows; they record that an order failed because a call to a backend service returned an error. The other templates are startup logs of \texttt{adservice} and transaction logs of \texttt{paymentservice}, with EOBs of 1.4\,\% and above. The trace block gives the average latency of each service in the window relative to its fault-free average (\S\ref{sec:design-input-profile}): \texttt{recommendationservice} rose to 5.5 times and first exceeded three times at 18:11, whereas \texttt{frontend} stayed at 0.8 times. The node thus has the rarest signal at the earliest time but neither the first position in the list nor the largest anomaly score, and the only evidence with an EOB of 0.0\,\% that names \texttt{frontend} is its two error logs.

\newcommand{\casefigbody}{%
  \centering
  \footnotesize
  \setlength{\fboxsep}{2pt}
  \begin{tabularx}{\linewidth}{@{}>{\raggedright\arraybackslash}X@{\hspace{1.5em}}>{\raggedright\arraybackslash}X@{}}
    \multicolumn{2}{@{}p{\linewidth}@{}}{\textbf{Case:} AIOps 2025 \texttt{qa\_229}. \quad \textbf{Ground truth:} \texttt{aiops-k8s-03} / \texttt{node fault} at 18:11. \quad \textbf{Observed in both responses:} \texttt{aiops-k8s-03} memory signal first crossed $z\ge3$ at 18:11 (fires in 0.00\,\% of normal windows); frontend log ``failed to complete the order: rpc error'' (0.0\,\%); recommendationservice latency $\times$5.5, frontend latency $\times$0.8.} \\[0.3ex]
    \toprule
    \textbf{(a) Off-the-shelf} & \textbf{(b) Ours} \\
    \midrule
    Alternatively, the \textbf{aiops-k8s-03} component has a node memory issue at 18:11:00, which is 0.00\% normal. [\ldots] \textbf{However, the logs are from the frontend.} [\ldots] So the component is frontend, reason is network attack, occurrence time 18:11:00.
    &
    However, the frontend's latency is only x0.8, which is lower than normal. \textbf{So maybe the frontend isn't the issue.} [\ldots] The aiops-k8s-03's first signal is infra\_node\_node\_memory\_MemAvailable\_bytes crossing z$\ge$3 at 18:11:00 with 0.00\% normal windows. [\ldots] \textbf{Maybe the node fault in aiops-k8s-03 caused the recommendationservice to have high latency.}
    \\[0.8ex]
    \caseanswer{red!10}{frontend / network attack}{wrong} & \caseanswer{green!12}{aiops-k8s-03 / node fault}{correct} \\
    \bottomrule
  \end{tabularx}
  \caption{\textbf{Root-cause selection before and after training on the same input.} One response per condition on the same input.}
  \label{fig:reasoning-case}
}
\placehereornext{\casefigbody}

The two responses perform the same analysis. Both examine the memory signal of the node, the error logs of \texttt{frontend}, and the latency ratios, and both note that the latency of \texttt{frontend} is 0.8 times its baseline. They diverge only at the decision: the off-the-shelf response lets the location of the error logs decide, reads the ratio of 0.8 as high, and selects \texttt{frontend}, the pattern of Finding 2 in which the final selection departs from the model's own analysis. The trained response weighs the same observations differently: the normal latency of \texttt{frontend} rules it out, and the latency of \texttt{recommendationservice} is attributed to the node fault. This is the decision that Biased-Decision Correction targets. Its pairs contrast continuations that share the analysis and differ in the selection (\S\ref{sec:design-pairs}), and the result is visible here: the off-the-shelf model names the node in the reasoning of seven of its eight responses but selects it in one, whereas the trained model names and selects it in all eight.

\section{Related Work}\label{sec:related}

\begin{autowrap}{table}{0.6\linewidth}
\centering
\newcommand{\cfull}{\tikz[baseline=-0.3em]\fill (0,0) circle (0.36em);}
\newcommand{\chalf}{\tikz[baseline=-0.3em]{\draw[line width=0.4pt] (0,0) circle (0.36em);\fill (0,-0.36em) arc[start angle=-90,end angle=90,radius=0.36em] -- cycle;}}
\newcommand{\cnone}{\tikz[baseline=-0.3em]\draw[line width=0.4pt] (0,0) circle (0.36em);}
\newcommand{\cna}{--}
\caption{\textbf{Properties of RCA approaches.}}
\label{tab:related}
\scriptsize
\setlength{\tabcolsep}{1pt}
\renewcommand{\arraystretch}{1.15}
\begin{tabular*}{\linewidth}{@{\extracolsep{\fill}}lccccc@{}}
\toprule
\textbf{Category} & \textbf{\makecell{Semantic\\reasoning}} & \textbf{\makecell{Normal\\behavior\\context}} & \textbf{\makecell{Full\\diagnosis}} & \textbf{\makecell{Decision-\\level\\training}} & \textbf{\makecell{Self-\\hosted\\LLM}} \\
\midrule
Statistical~\cite{fse26-torai, fse23-nezha, icse24-tracecontrast}         & \cnone & \cfull & \cnone & \cna   & \cna   \\
Prompted LLM~\cite{rcacopilot-eurosys24, openrca, arxiv25-microrcaagent}  & \cfull & \chalf & \cfull & \cna   & \chalf \\
Fine-tuned SLM~\cite{icse26-foundroot, tosem-thinkfl}                     & \cfull & \cnone & \chalf & \cnone & \cfull \\
Multi-agent SLM~\cite{ase-opsagent}                                       & \cfull & \cnone & \cfull & \cnone & \cfull \\
\makecell[l]{Reasoning preference\\learning~\cite{icml25-cdpo, arxiv24-stepdpo, emnlp24-selfexplore}} & \cna & \cna & \cna & \cfull & \cna \\
\midrule
\textbf{Ours}                                                          & \cfull & \cfull & \cfull & \cfull & \cfull \\
\bottomrule
\end{tabular*}

\par\smallskip
{\raggedright\scriptsize \cfull~full, \chalf~partial, \cnone~no support, and --~not applicable.\par}
\end{autowrap}

\noindent Table~\ref{tab:related} compares our system with existing approaches along five properties. Statistical methods reflect normal behavior but rank signals by statistical patterns rather than by what the signals mean, and they do not answer the fault type and time. LLM-based methods reason about what signals mean, but most compare them with the minutes around the incident window. With prompting alone, self-hosted models are far less accurate than commercial ones, and fine-tuned methods apply the training signal to whole responses. Preference learning for general reasoning applies the signal to the decision where reasoning diverges but does not address RCA. Our system combines these properties in a self-hosted 4B model, giving it how often each signal is flagged as anomalous without faults and training only the decision where its own reasoning diverges.

\textbf{Anomaly flagging in RCA.}
Existing methods take a baseline from the minutes around the incident window~\cite{ase-opsagent, icse26-foundroot} or from the fault-free period~\cite{fse26-torai, arxiv25-microrcaagent, icse24-tracecontrast, huo2023evlog} and give the model or the ranker how far each signal deviates from it. Nezha and SpecRCA also use how often something occurs without faults, but only to filter alerts or to adjust a score~\cite{fse23-nezha, icse26-specrca}. From the same fault-free period, our system counts how often each signal is flagged as anomalous without faults and gives this frequency to the model as the EOB next to the anomaly score. Even when the anomaly score alone cannot separate a routine deviation from a failure, the model can thus recognize a signal that is flagged as anomalous in the fault-free period as well. These baselines also use telemetry after the incident, from the minutes following the incident window~\cite{ase-opsagent, icse26-foundroot} to the normal interval after each incident~\cite{arxiv25-microrcaagent}, and our system likewise profiles fault-free windows on both sides of each incident.

\textbf{Learning RCA reasoning.}
The RCA training in \S\ref{sec:bg-finetune} applies the training signal to whole responses, and SpecRCA fine-tunes a 3B verifier on the responses of a teacher LLM while a large LLM synthesizes the final diagnosis~\cite{icse26-specrca}. Our system instead targets final selections that depart from evidence already in the input (Finding 2). It contrasts continuations that select different causes after the model's own common prefix and trains only this selection with DPO~\cite{neurips23-dpo}, without a teacher model or a larger model online.

\textbf{Preference construction for reasoning.}
Process supervision scores each reasoning step~\cite{iclr24-letsverify}, and Math-Shepherd~\cite{acl24-mathshepherd} and cDPO~\cite{icml25-cdpo} obtain such scores automatically by sampling continuations from the prefix of each step or token and measuring how often they reach the correct answer. Step-DPO~\cite{arxiv24-stepdpo} and Self-Explore~\cite{emnlp24-selfexplore} contrast correct and incorrect steps that share the prefix before the first error. Our system also samples continuations from prefixes of its own reasoning, but it sets the preference by the per-element answers of the RCA query, which also yields pairs between partially correct answers, and verifies each chosen response against the input. STaR gives the answer as a hint to obtain rationales for failed problems and fine-tunes on them without the hint~\cite{neurips22-star}. The privileged-context pairs of our system likewise drop the hint but pair the obtained continuation with the original incorrect one for DPO.

\section{Limitations and Threats to Validity}
\textbf{Fault-free profiles.}
As in prior work that uses telemetry from both before and after each incident~\cite{icse26-foundroot, arxiv25-microrcaagent}, our evaluation profiles fault-free windows on both sides of each incident, whereas a deployment would profile only earlier telemetry. RQ5 indicates that about half a day of fault-free windows near an incident suffices, and profiling only earlier windows remains future work. The profile may also lose relevance after the workload or service configuration changes, and a signal that is frequent during normal operation can still be related to a failure.

\textbf{Quality of preference pairs.}
Biased-Decision Correction depends on the training incidents and the quality of the preference pairs. A response with a correct final answer may still contain an incorrect causal explanation, and the reviewer LLM that checks it may misjudge it. The reviewer is a commercial model used only offline, and we did not evaluate a self-hosted model in this role.

\textbf{Implementation of baselines.}
The main threat to internal validity lies in the implementation of the baselines. KPIRoot+ is reimplemented from its paper due to the lack of public code, and KPIRoot is run from its public code with minor fixes. For all other methods, we use the official implementations, released checkpoints, or the authors' released results. Although we followed the described algorithms as closely as possible, potential discrepancies may exist.

\textbf{Scope of evaluation.}
The trajectory analysis targets Bank and Qwen3-4B, and the model comparison covers only the Qwen3 family and Gemma 4 E4B. The effects on other model families and on production environments that change over time, as well as on the investigation and recovery time of operators, require further validation.

\section{Conclusion}
In this work, we presented an SLM-based RCA system motivated by the gap between performing diagnostic analysis and selecting the root cause. 
Our empirical study showed that a model can examine candidates and propagation relationships yet still favor large anomaly scores in its final decision, while its input omits whether those anomalies are common during normal operation. 
Our system addresses these limitations by providing normal-behavior context and training the model on its own responses that reach different conclusions from the same initial analysis. With identical input, this targeted correction enables a 4B model to outperform an off-the-shelf 14B model. Using a single inference call per incident, our system achieves the highest exact-match accuracy on three of four evaluation datasets and ties with the best baseline on the fourth. These results highlight evidence interpretation and final root-cause selection as important targets for improving SLM-based RCA, alongside the diagnostic procedures that models follow.

\bibliographystyle{ACM-Reference-Format}
\bibliography{main}
\end{document}